\documentclass[ps4pdf,showpacs,aps,prd,twocolumn,nofootinbib,superscriptaddress]{revtex4}
\pdfoutput=1
\usepackage{amsmath,amssymb,amsbsy,graphics,graphicx,epsfig,array,latexsym,booktabs,docs}
\usepackage[colorlinks,hypertex]{hyperref}
\usepackage{ifthen,color,relsize,fancyvrb}
\hypersetup{pdfpagemode=FullScreen,urlcolor=black,citecolor=blue}

\begin{document}
\title{Searching Higgs in Noncommutative Electroweak Model at Photon-Photon Collider}

\author{Chien Yu Chen}
\email{d9522817@phys.nthu.edu.tw}
\affiliation{\small Department of Physics, National Tsing-Hua University, Hsinchu 300, Taiwan}

\begin{abstract}
We discuss the process of Higgs boson production in $\gamma\gamma$ collider on noncommutative spacetime and compare the results with large extra dimension in KK graviton channel. Summing all KK mode on IR brane, the affections are in the same order by comparing noncommutatve model prediction. This process is completely forbidden in standard model on unitarity condition and bosonic distribution. In noncommutative theory, the effect is induced by the coordinates noncommutable relation, $[x^{\mu}, x^{\nu}]$ = $i\theta^{\mu\nu}$. Due to the constant background strength tensor does not contain any conserved quantum number, hence, this effort is indicated into particle power spectrum. Particle mass spectrum is corrected by radiational and anisotropic surroundings. The process of $\gamma\gamma\to H^{0}H^{0}$ restricts the unitarity condition in noncommutative field theory. Under power law conservation condition, the neutral Higgs mass below gauge boson resonance will produce a accelerated phenomena as the central energy is higher than $Z_{0}$ gauge boson creation scale. The effects generated from the vast light Higgs particles descend the power rate energy distribution as far as the ambient is under a balance. The fractional rate on polarized polars are very small embedded into the unpolarized surroundings depend on background electric field couplings.
\end{abstract}

\maketitle

\section{Introduction}
In modern high energy collider physics, it is based on standard model prediction. The spontaneous symmetry breaking mechanics (SSB) used to generate particle mass from Higgs mechanics background methods. The $\gamma\gamma$ collider experiment~\cite{tesla} is playing an important role to generate a pure source of high energy gamma ray. The experiment parameters in wider energy range spectrum than $e^{+}e^{-}$ linear acceleration collider experiments in controlling electron polarization pole, $P_{e}$, and incoming laser circle polarization, $P_{laser}$ with the maximum energy range of 80$\%$ than $e^{+}e^{-}$ collider annihilation.

In this paper, we analyze $\gamma\gamma$$\to$$H_{0}H_{0}$ in noncommuative spacetime. Noncommutative geometry is to correct the spacetime on the background magnetic field. Lorentz is violated with a constant and uniformed background large scale magnetic and electric field. The background universe framework indicates an isolated direction to violate boost and rotation invariant under Lorentz transformation. In general Lorentz group $SO(1,3)$ symmetry, it is divided into O(1,1)$\otimes$SO(1,2) at an alternative choice of a different boost axis in background uniformed direction. The commutative relation generates a $\theta_{\mu\nu}$ deformed term~\cite{Seiberg:1999vs}, such as the strength tensor,
\begin{equation}
[\hat{x}^{\mu}, \hat{x}^{\nu}] = i\theta^{\mu\nu}.
\end{equation}

Noncommutative field theory contains an unknown mixture connection between IR and UV divergence. The nonlocal effects in above formula corrects the relation between particles under momentum space. The similar definition of UV/IR mixing denotes the diagrams condense the nontrivial overlap between different physics. Under unitarity constraint, the background electric field cannot take into account in higher than first order nonlocal perturbation~\cite{Ohl:2003zd}. In involving the nondiagonal components~\cite{Filk:1996dm} and discussing microcausality~\cite{Greenberg:2005jq} in quantum field properties, the restriction in causality requires that a constant background electric field will cause violating of quantum number distribution. Its violation indicates time arrow is not uniformed, in contrary to parity and charge asymmetry, it indicates that $U(1)$ and quantum field inner structure is not under a balance.

The processes of $\gamma\gamma\to$ Higgs do not appear parity violation in finial numerical analysis. The polarized fractional rate does not be changed in different incoming electron and photon pole. The consequence shows that the $\gamma\gamma\to$ massive neutral particle Higgs creation presents spontaneous breaking mechanics destroyed from break unitarity condition. The null state particle redefines its ground state vacuum via the road of inherent background duals. Violating unitarity condition is seemingly printed on $\gamma\gamma$ collider to Higgs particles producing events. By the way, it is different from the restriction on violating equilibrium in partition methods. It does not redefine the form of affected $Lagrangian$, $\mathcal{L}$ = $\mathcal{L}^{\dagger}$, but it combines the particle energy and momentum by unvanished nondiagonal components.

Otherwise, it is a $CPT$ conserved theory in four dimensional spacetime~\cite{Bazeia:2003vt}. The $CPT$ violated effect is to consider a 5D extension~\cite{Mocioiu:2001fx}. The parity violation produces from anisotropic geometry\footnote{Isotropic spacetime geometry conserves angular distribution and energy power spectrum by alternative coordinate translation. On the cosmic uniformed background field, we choose a frame to violate the coordinate translational invariance. In the background preferred frame, the effects induced by the remanent field is unobservable.}. Otherwise, due to noncommutative theory inherently contains dipole moment~\cite{Hinchliffe:2001im}, the charge violation is imposed on isotropic and homogeneous background universe. In U(1) model, the first order theta deformed term induces an anomaly electric dipole moment, $\bf{\mathcal{J}}\cdot\bf{\mathcal{E}}$, $CP$ violated effect is produced in assumed that background strength tensor is not a Lorentz symmetric product. On the background of Lorentz violation, parity is not conserved under the global symmetry. However, $CP$ asymmetry is, eventually, not a conserved quantity.

Charge is simultaneously violated in noncommutative electroweak theory, since triple photon coupling is produced naturally. $CP$ violation is unobservable on the process with triple gauge boson coupling. We cannot explore the time asymmetry effect from noncommutative spacetime structure with $CPT$ invariance. In standard model extension, scalar sector includes both $CP$-even, $H_{0}$ and $h_{0}$, and $CP$-odd, $A_{0}$, particles~\cite{Gunion:1984yn, Dawson:1989ws, Ellis:1988er, Muhlleitner:2001zv}. Those correspond to the terms $\vec{\epsilon}_{1}\cdot\vec{\epsilon}_{2})$ and $(\vec{\epsilon}_{1}\times\vec{\epsilon}_{2})\cdot\vec{k}_{\gamma}$ respectively. In noncommutative geometry, the momentum of $CP$-even, $H^{0}$, Higgs boson can couple to photon polarizations with, $\vec{p}_{1}\cdot\vec{\epsilon}_{1}$. However, $H^{0}$, Higgs boson with total angular momentum, $J$= 0, conserves parity asymmetry.

The process of $\gamma\gamma\to$ $H_{0}H_{0}$ is completely forbidden at tree level process, because the coupling, Z$\gamma\gamma$, cannot be predicted in standard model. It violates angular momentum conservation in loop order~\cite{Melic:2005am, Tureanu:2007zz}. Phenomenologically, it is produced under triple gauge boson condensation via Seiberg-Witten map~\cite{Tureanu:2007zz, Seiberg:1999vs}.
Therefore, significantly, zero total angular momentum, $J$= 0, contributes at event point. The mediate gauge boson does not interact with background field from angular power overlap. The distribution is dominated by the product of background field to particle polarization with total angular momentum, $J$= 0. It manifests that if we set electron polarization as same as laser photon, $p_{e}$ = 1 and $p_{l}$ = - 1, on the event point, the total photon luminosity will be maximum contributed. Contrastively, if we choose electron and incoming laser zero polarized pole, the contribution on the cross section only contains the minimal strength of event. We discuss the influence in comparing with each electron and laser photon polarization, $p_{e}$ and $p_{l}$. Finally, in photon power spectrum, high energy phenomena is poor to redistribute by exchanging the central energy range. At the central energy lower than 500 GeV, the change is dramatically independent on particle mass distribution.

\section{Noncommutative Higgs sector}
Higgs sector is a scalar field dominating particle mass, and contains all of the symmetry, $SU_{L}(N)\otimes SU_{R}(N)\to SU_{V}(N)$ in quantum field theory. Noncommutative theory is a perturbative theory coming from string theory without considering an exotic field. The general statistic
quantum field situation is still considered. Following the discussion of $CPT$ symmetry, the field average out of the equilibrium is disappeared in the restriction of $CP$ or $CPT$ conservation. Noncommutative theory generally describes a symmetry, $Pancar\check{e}$ transformation, on the alternative expansion of $\theta$ deformation. The boost effect does not produce a modification of particle statistical properties in global system. Scalar field moves on universe equilibrium, no $CPT$ violated effect is observed, such as baryogenesis and leptogenesis~\cite{Davoudiasl:2004gf}.

Otherwise, spin-1 vector potential, and spin-2 graviton are bosonic fields, its angular momentum polarized spectrum interacts with background field and generate different mode on the anisotropic spacetime background. In \cite{Buric:2006di, Rivelles:2002ez}, absorbing the $\theta$ deformed term into graviton field, and assuming that the geometric fluctuations is composed by the tensor field, $\theta_{\mu\nu}$,
\begin{equation}\begin{split}\label{eq1}
h^{\mu\nu} &= \theta^{\mu\alpha}F_{\alpha}^{~\nu} + \theta^{\nu\alpha}F_{\alpha}^{~\mu} + \frac{1}{2}\eta^{\mu\nu}\theta^{\alpha\beta}F_{\alpha\beta},\\
h &= 0,\\
\end{split}\end{equation}
and \cite{Brandt:2006ua} deforms spacetime coordinate on the module expansion, $\hat{x}^{\mu} = x^{\mu} + \theta^{\mu\nu}\partial_{\nu}f(x)$. In addition, some papers consider nonsymmetric matrix to model a noncommutative geometry~\cite{Moffat:1995fc}, and discuss a nonsymmetric graviton on annihilation process~\cite{Kersting:2003ea}.

Modifying background polarization is equal to redefine the angular power spectrum and particle mass distribution. Mapping gauge boson to noncommutative spacetime by Seiberg-Witten map
\begin{equation}\label{eq2}
\widehat{V}_{\mu} = V_{\mu} + \frac{1}{4}\theta^{\alpha\beta}\big{\{}\partial_{\alpha}V_{\mu} + F_{\alpha\mu}, V_{\beta}\big{\}} + \mathcal{O}(\theta^{2}),
\end{equation}
where noncommutative deformed vector potential is constructed by standard model field and $\theta_{\mu\nu}$ deformed terms. In general gauge potential on the non-abelian gauge, U(1) quantum number is constrained by the relation, $Q$ = $T_{3}$ + $Y$, $Y$ is hypercharge and $T_{3}$ is nonabelian traceless eigenvalues. The general expression is
\begin{equation}\label{eq3}
V_{\mu} =g'A_{\mu}Y + g\sum^{3}_{i=1}B^{i}_{\mu}T^{i}_{L} + g_{S}\sum^{8}_{i=1}G^{i}_{\mu}T^{i}_{S},
\end{equation}
extended by U(1) and non-abelian gauge group generators. Splitting U(1) gauge potential into two part, contains both hypercharge $Y$ = $\frac{1}{2}$ with different gauge couplings. Under unitarity gauge, Higgs field is rotated to
\begin{eqnarray}\label{eq4}
\widehat\Phi = \left(\begin{array}{c} \widehat\Phi^{+} \\ \widehat\Phi^{0}
\end{array}\right),
\end{eqnarray}
the charged Higgs is forbidden by choosing unitarity transformation. It preserves the neutral Higgs on vacuum and a constant $vev$ from SSB breaking scalar potential.

In NC electroweak model~\cite{Melic:2005am}, Higgs field is under the fundamental and anti-fundamental transformation, due to Yukawa term couples to the left-handed fermion and right-handed fermion under different representation of $SU_{L}(2)$ and $U_{Y}(1)$. The expression on noncommutative deformed representation is described as
\begin{equation}\label{eq5}
\begin{split}
S_{Higgs} = \int d^{4}\bigg{\{}\frac{1}{2}(\widehat{D}_{\mu}\widehat{\Phi})^{\dagger}\star&(\widehat{D}^{\mu}\widehat{\Phi}) - \frac{\mu^{2}}{2!}\widehat{\Phi}^{\dagger}\star\widehat{\Phi}\\
 &-\frac{\lambda}{4!}\widehat{\Phi}^{\dagger}\star\widehat{\Phi}\star\widehat{\Phi}^{\dagger}\star\widehat{\Phi}\bigg{\}},\\
\end{split}
\end{equation}
the general expression of scalar field and its covariant derivative are
\begin{equation}\label{eq6}
\begin{split}
\widehat{\Phi} &= \Phi +\frac{1}{2}\theta^{\alpha\beta}V^{m}_{\beta}D^{\Phi}_{\alpha}\Phi,\quad D^{\Phi}_{\alpha} = \partial_{\alpha} - iV^{\Phi}_{\alpha},\\
\widehat{D}_{\mu}\widehat{\Phi} &= \partial_{\mu}\widehat{\Phi}-i[\widehat{V}^{-\frac{1}{2}}_{\mu}\star\widehat{\Phi} - \widehat{\Phi}\star\widehat{V}^{-1}_{\mu}] - i\widehat{V}^{SU(2)}_{\mu}\star\widehat{\Phi},\\
\end{split}
\end{equation}
where $\widehat{V}^{-\frac{1}{2}}_{\mu}$ and $\widehat{V}^{-1}_{\mu}$ are vector fields, Eq.($\hyperref[eq2]{2}$), with the hypercharge are $- \frac{1}{2}$, and $- 1$. The general expression of gauge potential splits gauge boson into hypercharge $Y$ = $- \frac{1}{2}$, $-1$, and non-abelian terms.

Vector fields overlap with Higgs field range generates a mass distribution. On the other hand, particle mass is dominated with the area coupled by Higgs field. Photon couples to Higgs field generates a range producing an overlap with scalar vacuum.
\begin{widetext}
\begin{equation}\label{eq7}
\begin{split}
\widehat{V}^{\Phi}_{\mu} &= \widehat{V}^{-\frac{1}{2}} - \widehat{V}^{-1} + \widehat{V}^{SU(2)}_{\mu}
= \left(\begin{array}{@{}cc@{}}
   e\widehat{A}_{\mu} + e\cot2\theta_{W}\widehat{Z}^{0}_{\mu} & \frac{e}{\sqrt{2}\sin\theta_{W}}\widehat{W}^{+}_{\mu}\\
   \frac{e}{\sqrt{2}\sin\theta_{W}}\widehat{W}^{-}_{\mu} & -\frac{e}{\sin2\theta_{W}}\widehat{Z}^{0}_{\mu}\\
   \end{array}\right)\\
\widehat{V}^{m}_{\mu} &= \widehat{V}^{-\frac{1}{2}} + \widehat{V}^{-1} + \widehat{V}^{SU(2)}_{\mu}
= \left(\begin{array}{@{}cc@{}}
   -e\widehat{A}_{\mu} + \frac{e}{2}(3\tan\theta_{W} + \cot\theta_{W})\widehat{Z}^{0}_{\mu} & \frac{e}{\sqrt{2}\sin\theta_{W}}\widehat{W}^{+}_{\mu}\\
   \frac{e}{\sqrt{2}\sin\theta_{W}}\widehat{W}^{-}_{\mu} & -e\widehat{A}_{\mu} + \frac{e}{2}(3\tan\theta_{W} - \cot\theta_{W})\widehat{Z}^{0}_{\mu}\\
   \end{array}\right).\\
\end{split}
\end{equation}
\end{widetext}
The mixture part devotes on the coupling of one gauge boson to Higgs field. On this process, we extract the coupling of photon and $Z^{0}$ to two Higgs. Testing photon-photon collider experiments, we get a consistent signature in cosmology background field direction. Following~\cite{Melic:2005am}, we rewrite the Higgs sector after correcting the guage potential. Splitting gauge field into standard and vector, including $Y$=$-1$ and $Y$=$- \frac{1}{2}$ respectively.

\begin{figure}[http]
   \centering
   \includegraphics[width=3.5in]{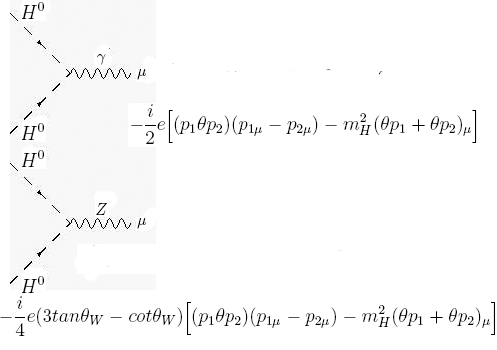}
   \caption{The couplings of neutral Higgs to two chargeless gauge bosons. These couplings are completely forbidden by standard model prediction.}
   \label{fig:FR}
\end{figure}

In order to put a constraint on U(1) gauge, $Q$ = $T_{3}$ + $Y$, the coupling constant on $A_{\mu}$ and $Z^{0}_{\mu}$ gauge field is modified. Photon and $Z^{0}_{\mu}$ U(1) gauge boson are the same kind fields on the second term in Eq.($\hyperref[eq7]{7}$). $Z^{0}_{\mu}$ is a massive particle, photon and neutral $Z^{0}_{\mu}$ gauge boson with the same generator, it induces a clue of the split between photon with $Z^{0}_{\mu}$ fields by translating the splitting to different coupling constants. The phenomena presents in enveloping algebra to expand gauge sector. In noncommutative spacetime structure, the mixture term induces a forbidden couplings, Fig.($\ref{fig:FR}$), and modifies the relation between neutral Higgs and U(1) gauge boson angular distribution on 4D brane.

In Eq.($\hyperref[eq7]{7}$), $Z^{0}_{\mu}$ gauge boson mixes with photon, the general Higgs sector is
\begin{equation}\label{eq8}
\begin{split}
S_{Higgs} &= \frac{1}{2}\int d^{4}x\big{[}(D^{\Phi}_{\mu}\Phi)^{\dagger}(D^{\mu}_{\Phi}\Phi) - \mu^{2}|\Phi|^{2} - \lambda|\Phi|^{4}\big{]}\\
     &\quad +\frac{\theta^{\alpha\beta}}{4}\int d^{4}x\Phi^{\dagger}\bigg{\{}U_{\alpha\beta} + U^{\dagger}_{\alpha\beta} + \frac{\mu^{2}}{2}F^{m}_{\alpha\beta}\\
     &\quad - i\frac{\lambda}{12}\Phi\big{[}(\partial_{\alpha}\Phi)^{\dagger}D^{m}_{\beta} + i\Phi^{\dagger}V^{m}_{\alpha}D^{\Phi}_{\beta}\big{]}\\
     &\quad - i\frac{\lambda}{12}\Phi\big{[}(D^{m}_{\alpha}\Phi)^{\dagger}\partial_{\beta} - i(D^{\Phi}_{\alpha}\Phi)^{\dagger}V^{m}_{\beta}\big{]}\\
     &\quad - i\frac{\lambda}{12}\Phi\Phi^{\dagger}\big{[}V^{\Phi}_{\alpha}V^{m}_{\beta} + V^{m}_{\beta}V^{\Phi}_{\alpha}\big{]}\bigg{\}}\Phi,\\
\end{split}
\end{equation}
where $F^{m}_{\alpha\beta}$ is $Maxwell$ tensor with $V^{m}_{\mu}$, Eq.(8), and the tensor field, $U_{\alpha\beta}$,
\begin{equation}\label{eq9}
\begin{split}
&U_{\alpha\beta} =\\
&(\overleftarrow{\partial}^{\mu} + iV^{\Phi\mu})\bigg{[}-(\partial_{\mu}V^{m}_{\alpha})\partial_{\beta} - V^{m}_{\alpha}\partial_{\mu}\partial_{\beta} + (\partial_{\alpha}V^{m}_{\mu})\partial_{\beta}\\
&+ iV^{m}_{\alpha}(\partial_{\mu}V^{\Phi}_{\beta}) + iV^{m}_{\alpha}V^{\Phi}_{\beta}\partial_{\mu} + iV^{\Phi}_{\mu}V^{m}_{\alpha}\partial_{\beta} + i(\partial_{\mu}V^{m}_{\alpha})V^{\Phi}_{\beta}\\
&+ \frac{i}{2}\big{\{}V^{\Phi}_{\alpha}, \partial_{\beta}V^{\Phi}_{\mu} + F^{\Phi}_{\beta\mu}\big{\}} + V^{\Phi}_{\mu}V^{m}_{\alpha}V^{\Phi}_{\beta}\bigg{]},\\
\end{split}
\end{equation}
where $F^{\Phi}_{\beta\mu}$ is made of $V^{\Phi}_{\mu}$ potential, Eq.(8). Using Eq.($\hyperref[eq6]{6}$) and Eq.($\hyperref[eq7]{7}$) in Higgs sector, we get the feymann rules, Fig.($\ref{fig:FR}$), on complete $\theta_{\mu\nu}$ deformed coupling constant. The quantity associates with the background field direction and the incoming particle energy.

On the background surroundings, the deformed magnetic and electric field produce a discrete patch to modify quantum field theory. The interaction with other fields from coupling to background $\theta_{\mu\nu}$ perturbation, and the efforts are coming from the overlap with large scale magnetic field. Therefore, in this paper, we focus on the discussion with the coupling between Higgs and U(1) gauge boson in considering the process of high energy gamma ray to two neutral Higgs particles.

\section{Photon Collider Experiments}
On the next linear collider, it is in order to search Higgs particle, superpartner, and towards to explore the Planck scale physics at TeV energy scale. Large Hardron Collider (LHC), International Large Collider(ILC), and photon collider experiment (TESLA), etc.. In this paper we work in probing Higgs particle at pure source of photon-photon collider experiments data analysis. The photon collider at TESLA is using two observable
technology methods to generate pure gamma ray. First is $\gamma\gamma$ technique methods, in considering polarized electric beam, $p_{e}$, interfered with polarized high energy gamma ray, $p_{l}$~\cite{tesla} at electron bunch point. The other is $\gamma e$ photon collider into electron target. The total incoming photon luminosity, $L_{\gamma\gamma}$, is around $\frac{1}{3}L_{e^{+}e^{-}}$ electron beam production. The background of photon collider made of the polarized electron beam at electron bunch point. We rearrange incoming electron beam in setting the polarized parameter $p_{e}$ = 0 or 1 for unpolarized and polarized channel, and $p_{\gamma}$ = -1, 0, 1 for interacted channel.

Different polarized electron and laser photon beam produces a different incident source from backreact scattering effects. The energy scale at 800 GeV, photon luminosity approaches to 1.7 ($cm^{-2}\cdot s^{-1}$) magnitudes under electron luminosity around 5.8 unit. In considering the polarized electron beam and photon beam, we have to take photon polarization distribution function, $\xi(x)$, into account. The result is under multiplying fractional luminosity and integrating the energy rate, x and y. The central energy, $\sqrt{s}$ = $E_{CM}$,  $E_{CM}$ = $xE_{1}$ + $yE_{2}$, constrains laser energy range in x + y =1 on the incident point under energy conservation condition. The integration range is decided by creating mediate particle and the kinetic momentum of finial state particle. The final particle kinetic energy has to be positive, and the fractional energy range truncates on the branch $\frac{M_{Z}}{\sqrt{s}}$.

The spin-1 particles contain 3 polarized state, spin up, spin-0, and spin down. The positive channel is canceled out by opposite sign, due to helicity conservation condition. The zero channel is parity conserved without exotic background influence. The corresponded zero polarized cross section is
\begin{equation}\begin{split}\label{eq10}
\sigma_{C} &= \frac{v}{s}\bigg{(}\frac{\alpha}{2}\bigg{)}^{2}\int^{xm} d\Omega|M_{J_{z}=0}|^{2}\\
|M_{J_{z}=0}|^{2} &= \frac{|M_{++}|^{2} + |M_{--}|^{2}}{2},\\
\end{split}\end{equation}
where $v$ is the velocity of final state particle, $\alpha$ is fine structure constant, and $x_{m}\approx 2( + \sqrt{2})$ is maximum energy fractional rate in photon collider. The luminosity amplitude in photon collider background,
\begin{eqnarray}
\sigma_{L} = \int dxdy f(x)f(y)\bigg{(}\frac{1+\xi(x)\xi(y)}{2}\bigg{)}\sigma_{C},
\end{eqnarray}
integrating out the fractional energy parameter x and y of incident laser photon with the velocity, $v$, of outgoing massive particle. Under the condition of parity conservation, each incoming laser photon cross section is consistent with different final helicity state where the "+" and "-" denote different circle polarization. In Ref(\cite{Cheung:2005pg}), it considers radion particle field in Randall-Sundrum model at photon collider experiments.

In next section we briefly introduce the Higgs generation process on large extra dimension model. The devotion is very small by comparing noncommutative process with the discussed polarized luminosity function~\cite{Cheung:2005pg},
\begin{equation}\label{eq11}
f(x) = \frac{dL_{\gamma\gamma}}{dz}\frac{1}{L_{geom}},
\end{equation}
where the background luminosity,
\begin{equation}\label{eq12}
\begin{split}
L_{geom} &= \bigg{[}\big(1-\frac{4}{z}-\frac{8}{z^{2}}\big)ln(z+1)+\frac{1}{2}+\frac{8}{z}-\frac{1}{z(z+1)^{2}}\bigg{]}\\
                 &+p_{e}p_{l}\bigg{[}\big(1+\frac{2}{z}\big)ln(z+1)-\frac{5}{2}+\frac {1}{z+1}-\frac{1}{2(z+1)^{2}}\bigg{]},\\
\end{split}
\end{equation}
and double photons luminosity,
\begin{equation}\label{eq13}
\begin{split}
\frac{dL_{\gamma\gamma}}{dz} &= \frac{1}{1-x}+(1-x)-4r(1-r)\\
                         &\quad -p_{e}p_{l}rz(2r-1)(2-x).\\
\end{split}
\end{equation}
The polarization of initial state photon associates with the electron and photon polarized beam, $p_{e}$ and $p_{l}$,
\begin{equation}\label{eq14}
\begin{split}
\xi(x) &= \frac{1}{L_{geom}}\bigg{\{}p_{e}\big[\frac{x}{1-x}+x(2r-1)^{2}\big]\\
                                            &-p_{l}(2r-1)\big[1-x+\frac{1}{1-x}\big]\bigg{\}}.\\
\end{split}
\end{equation}
The photon polarized strength depends on incoming electron polarization and laser photon polarization. It generates the transverse and linear photon state by controlling the incoming electron and laser beams.

\section{Large Extra Dimension model}
Extra dimension is to extend ours spacetime in $4+n$ dimensions under $SO(1,n+3)$ group without modifying time component. There are two aspects to introduce extra dimension model. First, considering the polynomial expansion, the coordinate adds a fifth dimensional coordinate variable, $x^{n}$ = $\frac{\vec{n}\cdot\vec{y}}{R}$~\cite{Antoniadis:1998ig, Kundu:2008fj}. In general, the geometry in extra dimension framework is written as, $g^{\hat{\mu}\hat{\nu}}$. The index $\hat{\mu}$ and $\hat{\nu}$ denote $n+4$ dimensions, from 4 to extra 4+n. The variable y is the fifth extra coordinate space within particle field projects to bulk in the radius $R$, and the extra extended spacetime, y, with KK modes expansion. Second, considering the two p-brane world-sheet~\cite{Randall:1999ee}, p = 3, on the warped bulk surroundings, one is IR visible brane, and another is UV invisible brane. The particle field is separated into fifth dimension and origin four dimensional field. Summing all influences on warped extra dimension by integrating out the 5D component y gets an exotic mixing between different particle field generations, such as flavor changing neutral current and neutrino mass~\cite{Dvali:1999cn}.

In general consideration, particle field is survived in 4D brane. In large extra dimension case, the extra fifth dimension produces a closed loop on the bulk space, from a 4D point to bulk field and coming back to the same point. This circle phenomenon is enclosed in a box space containing bulk wave function. In this box, each wave function on the ending point is disappeared, the bulk space produces a static wave. Each nodal point regards as a cutting dot. The stored energy~\cite{Cheng:2004ez} and produced pressure~\cite{Frank:2008dt} on the box surface is coming from the wave function dispersion into extra dimensions. Summing over the devotion, particle field is sinuously corrected.
The mass on bulk radius is deduced from the d$\tilde{O}$Alembert operator. In graviton case, $\square_{4+n}\big{(}\hat{h}^{\hat{\mu}\hat{\nu}} - \frac{1}{2}\eta_{\hat{\mu}\hat{\nu}}\hat{h}\big{)}$ =$\hat{h}^{\hat{\mu}}_{~\hat{\mu}}$ = 0, are produced onto graviton field. Then, after redefining, extracting the physical fields, the bulk mass term is produced from $m^{2}_{\vec{n}}$ = $\frac{4\pi^{2}\vec{n}^{2}}{R^{2}}$. On 4D brane, $m_{\vec{n} = 0}$ = 0, graviton is a massless field.

\begin{figure}[htbp]
   \centering
   \includegraphics[width = 1in]{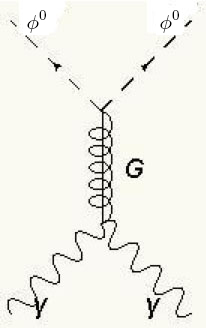} 
   \caption{$\gamma\gamma\to Grav.\to\phi^{0}\phi^{0}$}
   \label{fig:EDG}
\end{figure}
\begin{table}
\caption{Large extra dimension data analysis at $E_{CM}$ = 800 GeV, and on the string scale $M_{S}$ = 1 TeV}
\label{tab:Lar}
\begin{ruledtabular}
\begin{tabular}{cccc}
$\sigma^{n}_{pol.}/\sigma^{n}_{Unpol.}$&$p_{e}$=1,$p_{l}$=-1&$p_{e}$=1,$p_{l}$=0&$p_{e}$=1,$p_{l}$=1\\ \hline
n = 0                     &  0.112538                 & 0.055313                  & 0.0207263                \\
n = 2                     & 0.465354                  & 0.224496                  & 0.0895319                \\
n $>$ 2                & 0.0620941/(n-2)                 & 0.0319583/(n-2)                 & 0.0152298/(n-2)                \\
\end{tabular}
\end{ruledtabular}
\footnotetext[1]{The Higgs boson mass $m_{\phi}$ at 150 GeV.}
\end{table}

Focusing on the graviton field~\cite{Han:1998sg}, extending matric into extra dimensionsl indexes, graviton field is a geodesic fluctuations products. On fifth dimensional spacetime, the exotic space-like vector on fifth graviton field is merged, and other scalar components are also imposed in the geometric inflation. In considering the gauge condition, assuming no background field in gravitational radiation range, graviton field is re-composed into the other physical situations. On this model~\cite{Han:1998sg}, all the matter field is restricted in 4D brane, only graviton field can project to fifth dimensional spacetime. We discuss the phenomenon into probing Higgs field at photon-photon collider experiments. In considering $Casimir$ effect, the stored bulk energy is a curious point to take the incident into account. Using the feymann rule in Ref.(\cite{Han:1998sg}), the total cross section in even dimension is listed in Tab.I.

The amplitude is written as
\begin{equation}\begin{split}\label{eq16}
M_{\gamma\gamma\to G\to\phi\phi} = - iD_{n}\bigg{(}\frac{k}{2}\bigg{)}\epsilon^{\mu}_{1}\epsilon^{\nu}_{2}\bigg{(}\frac{s}{2}&c_{\mu\nu, \alpha\rho} + D_{\mu\nu, \alpha\rho}\bigg{)}B^{\alpha\rho, \beta\sigma}\\
&(m^{2}_{\phi}\eta_{\beta\sigma} - c_{\beta\sigma, \xi\eta}k^{\xi}_{1}k^{\eta}_{2}),\\
\end{split}\end{equation}
where the tensor field $C_{\rho\sigma, \xi\eta}$ and $D_{\mu\nu, \alpha\rho}$ is the coupling tensor. The $\frac{1}{2}B^{\alpha\rho, \beta\sigma}$ is graviton spin polarization sum in bulk spacetime with mass $m_{\vec{n}}$ and the $D_{n}$ is graviton projector operator.\\
In n = 0 case,
\begin{equation}
D_{0}~=~\frac{i}{s},
\end{equation}
In n = 2 case,
\begin{equation}
D_{2}~=~-\frac{i}{4\pi}R^{2}Log(\frac{M^{2}_{S}}{s}),
\end{equation}
In n$\geq$2 case,
\begin{equation}
D_{\geq 2}~=~-\frac{2i}{(n-2)\Gamma(\frac{n}{2})}\frac{R^{n}M^{n-2}_{S}}{(4\pi)^{\frac{n}{2}}}.
\end{equation}
The polarization rate are listed in Tab.I with each n numbers. Due to extra dimension, $y$, works on odd dimensional bulk space. There is an anomaly problem generated on the odd extra dimension numbers, n. In Tab.I, n = 2 extra mode, the polarized rate is maximum contributed on the Higgs creation process.

The radius $R$, cosmological constant and string effective scale $M_{S}$ are redefined into the restriction $\frac{1}{G_{N}}\approx M^{n+2}_{S}R^{n}$. In the $nth$ extra dimension devotion, the relation between gravitational constant $\kappa$ and $R^{n}$ are inside into a range of $16\pi(4\pi)^{\frac{n}{2}}\Gamma(\frac{n}{2})M^{-(n+2)}_{S}$. We take the mainly domain on the assumption of $M_{S}>>s$. The bulk field in extra dimensional box $\frac{2}{\Gamma(\frac{n+4}{2})}\pi^{\frac{n+4}{2}}R^{n}$. In exculding $Carsmir$ effect, the single graviton mode is similar order magnitude in considering the string effective scale, $M_{S}$.

\section{The process of $\gamma\gamma\to$ Higgs}
The nature source of high energy gamma ray collider is coming from sky. The process of $\gamma\gamma$ collider is a frequent incident in astroparticle physics. We test these phenomena from linear high energy $\gamma\gamma$ collider experiments, and model the cosmological incidents on large scale background. On the noncommutative spacetime, the background magnetic field and electric field will interact with particle spin and electric charge via $\theta_{\mu\nu}$ connection. From Eq.($\hyperref[eq11]{11}$), we clearly understand the background field how to produce with the initial state photon polarized vector and interact with final charged fields. We analyze the particle power spectrum in $\gamma\gamma$ collider experiments. This dramatic field interaction induces a sensitive effects associated with the $Planck$ scale $\Lambda_{c}$ and the background direction on the finial physical result.

\begin{table*}
\caption{Quantum numbers of Higgs and Gauge Bosons}
\label{tab:QN}
\begin{ruledtabular}
\begin{tabular}{cccc}
& \textbf{$J^{PC}$} & &\textbf{$J^{P}$}\\ \hline
\multicolumn{4}{c}{\textcolor{blue}{\textbf{When $C$ and $P$ are separately conserved}}}\\
$\gamma$ & $1^{--}$   & $W^{\pm}$ & $1^{-}$\\
             $Z$ & $1^{--}$   & $H^{\pm}$ & $0^{+}$ \\
     $H^{0}$ & $0^{++}$ &                     & \\
      $h^{0}$ & $0^{++}$ &                     & \\
      $A^{0}$ & $0^{+-}$  &                     & \\
                     &                  &                      &\\
\multicolumn{4}{c}{\textcolor{blue}{\textbf{When $C$ and $P$ are violated but $CP$ still conserved}}}\\
$\gamma$ & $1^{--}$    & $W^{\pm}$ & $1^{-}$, $1^{+}$\\
$Z$             & $1^{--}$, $0^{++}$    & $H^{\pm}$  & $0^{+}$, $0^{-}$ \\
$H^{0}$      & $0^{++}$, $0^{--}$  &                     & \\
$h^{0}$      & $0^{++}$, $0^{--}$  &                      & \\
$A^{0}$      &$0^{+-}$,  $0^{-+}$  &                      & \\
\end{tabular}
\end{ruledtabular}
\footnotetext[1]{The table in $pages~198$, $The~Higgs~Hunter's~Guide$}
\end{table*}
Tab.II lists the quantum numbers at Higgs particle internal symmetry. We use this discussed particle properties on the $\gamma\gamma\to$ Higgs process. Otherwise, Tab.III concerns the model restriction on the standard 2HDM and the supersymmetry 2HDM. The particle mass spectrum is defined by the two vacuum $v_{1}$ and $v_{2}$, the axion mass is a free parameter on the model building, another is the vacuum fractional rate of $\tan\beta$.

\begin{table*}
\caption{The experiments bounds on $\tan\beta$ and scalar mass $m_{H^{0}}$ and $m_{H^{\pm}}$}
\label{tab:EX}
\begin{ruledtabular}
\begin{tabular}{cc}
           &\textbf{$\tan\beta$} and $m_{H^{0}}$\\ \hline
ACHARD 03C & $m_{H^{0}}>108.1~GeV$. For $B(H^{0}\to\gamma\gamma)$=1, $m_{H^{0}}>114~GeV$.\\
ABBIENDI 02D & 4$<m_{H^{0}}<12~GeV$\\
           & For $B(H^{0}\to\gamma\gamma)$=1, $m_{H^{0}}>117~GeV$.\\
\qquad\qquad\quad~99E& $m_{H^{0}}$ = $m_{A^{0}}$ in general 2DHM.\\
ACCIARRA 00s& The limits on $\Gamma(H^{0}\to\gamma\gamma)\cdot$B($H^{0}\to\gamma\gamma$ or $b\bar{b}$) for $m_{H^{0}}>98~GeV$\\
           & For B($H^{0}\to\gamma\gamma$) = 1, $m_{H^{0}}>98~GeV$.\\
BARATE 00L& For B($H^{0}\to\gamma\gamma$) = 1, $m_{H^{0}}>109~GeV$.\\
ABBOTT 99B& Limits in the range of $\sigma(H^{0}+Z/W)\cdot$B($H^{0}\to\gamma\gamma$) = 0.80-0.34 pb are obtained in $m_{H^{0}}$ = 65-150 GeV.\\
ALEXANDER 96H& B($Z\to H^{0}\gamma$)$\times$B($H^{0}\to q\bar{q}$)$<$ 1-4$\times 10^{-5}$ 95$\%$ CL\\
             & and B($Z\to H^{0}\gamma$)$\times$B($H^{0}\to b\bar{b}$)$<$0.7-2$\times 10^{-5}$ 95$\%$ CL in 20$<m_{H^{0}}<80$ GeV.\\ \hline
           &\textbf{$\tan\beta$} and $m_{H^{\pm}}$\\ \hline
ABULENCIA 06E& Within MSSM, the range $\tan\beta<$1 or $>$30 in $m_{H^{\pm}}$ = 80-160 GeV.\\
ABBIENDI 03& $m_{H^{\pm}}>$1.28$\tan\beta$ GeV 95$\%$ CL in type II.\\
\qquad\qquad\qquad~01Q & $\tan\beta <0.53~m_{H^{\pm}}~GeV^{-1}$ 95$\%~CL$ in type II.\\
ABAZOV 02B & $\tan\beta >32.0$ excluded at 95$\%$CL $m_{H^{\pm}}$ = 75 GeV.\\
           &The excluded mass range extends to over 140 GeV for $\tan\beta>$100.\\
BARATE 01E & $\tan\beta <0.40~m_{H^{\pm}}~GeV^{-1}~90\%~CL$ in type II and $<~0.49~m_{H^{\pm}}~GeV^{-1}~90\%~CL$.\\
AFFOLDER 001& For $\tan\beta >100$, $m_{H^{\pm}}<120 GeV$. If B($t\to bH^{+}$)$\gtrsim$0.6, $m_{H^{\pm}}$ up to 160 GeV.\\
ABBOTT 99E& $\tan\beta\lesssim$1, $120\gtrsim m_{H^{\pm}}>$50 GeV. For $\tan\beta\gtrsim$40, 160$\gtrsim m_{H^{\pm}}>$50 GeV.\\
ACKERSTAFF 99D& For 2DHM, only on doublet couples to lepton, $m_{H^{\pm}}>0.97\tan\beta$ GeV 95$\%$ CL.\\
ACCIARRI 97F& $m_{H^{\pm}}>2.6\tan\beta$ GeV 90$\%$ CL from excluding B$\to\tau\nu_{\tau}$ branching ratio.\\
AMMAR 97B& In lowr limits, $m_{H^{\pm}}>0.97\tan\beta$ GeV 90$\%$CL.\\
STAHL 97& $m_{H^{\pm}}>1.5\tan\beta$ GeV 90$\%$ CL.\\
ALAM 95& The limits $m_{H^{\pm}}>$244+63/$\tan\beta^{1.3}$ in the 2DHM. Light supersymmetry can invalidate this bounds.\\
BUSKULIC 95& $m_{H^{\pm}}>$1.9$\tan\beta$ GeV 90$\%$CL in type II.\\
\end{tabular}
\end{ruledtabular}
\footnotetext[1]{The data is revealed in $Particle~Data~Group$.}
\end{table*}

\subsection{$\gamma\gamma\to H^{0}H^{0}$}
In this section, we introduce the process and analyze the properties of polarized and unpolarized beam. Comparing the probability distribution and particle mass spectrum in the different fields. The total cross section is maximum rearranged by background magnetic and electric field perpendicular to the incoming particle direction. The ratio of polarized and unpolarized beam under different polarization electron and incoming laser photon, we set the Higgs mass, $m_{H^{0}}$ = 150 GeV, $m_{H^{\pm}}$ = 120 GeV, and $m_{A^{0}}$ = 100 GeV. The central energy up to
800 GeV with the scale of $\Lambda_{c}$ = 1 TeV and the triple gauge boson coupling $K_{Z\gamma\gamma}$ = - 0.2, $K_{\gamma\gamma\gamma}$ = - 0.3. There is a minimum devotion on the central energy around the range 100 GeV $\sim$ 500 GeV.

\begin{figure}[http] 
   \includegraphics[width=1in]{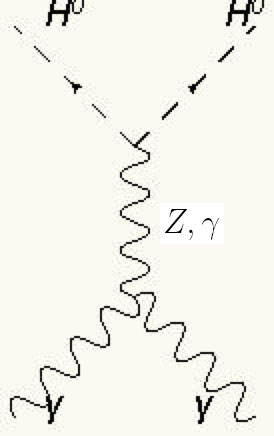}
   \caption{$\gamma\gamma\to z\gamma\to H_{0}H_{0}$.}
   \label{fig:gh}
\end{figure}

The $\gamma\gamma$ collider technology can be used to simulate cosmology high energy gamma ray annihilation process. Background high energy frequency is produced by photon gas entropy. The gradient distribution is a source to modify angular power spectrum. Background angular distribution on CMB experiments shows us that our universe is not perfect isotropic. There is a preferred direction imposed on background field and produces a temperature of non-uniformed distribution. On this process, we get an incident to generate scalar particle from high energy gamma ray annihilation. Scalar particle on the cosmological potential $V(\phi)$ is a source to induce inflation effects. On magnetic field and electric field background in considering the stored energy, noncommutative electroweak model generates a theoretic clue to drive coupling $V-H^{0}-H^{0}$ and produces triple gauge boson couplings. The simulation results in a defined source higher than $e^{+}e^{-}$ linear collider at almost one order magnitude. It is experiment data to consider Higgs particle production and the existence opportunity of inflaton particle.

Using Feymann rule, Fig.($\ref{fig:FR}$), the amplitude is written as a process contains $\gamma$ and Z resonances,
\begin{eqnarray}\label{eq16}
\begin{split}
M_{\gamma\gamma\to H_{0}H_{0}} =& -\frac{i}{s}G_{\gamma\gamma\gamma}G_{H_{0}H_{0}\gamma}\\
                                                                   & - \frac{i}{s - m^{2}_{H_{0}}-im_{Z}\Gamma_{Z}}G_{\gamma\gamma Z}G_{H_{0}H_{0}Z},
\end{split}
\end{eqnarray}
where
\begin{equation}\begin{split}\label{eq17}
G_{\gamma\gamma\gamma}G_{H_{0}H_{0}\gamma} =& G_{\gamma\gamma\gamma_{L}}G_{H_{0}H_{0}\gamma_{R}} + G_{\gamma\gamma\gamma_{R}}G_{H_{0}H_{0}\gamma_{L}},\\
G_{\gamma\gamma Z}G_{H_{0}H_{0}Z} =& G_{\gamma\gamma Z_{L}}G_{H_{0}H_{0} Z_{R}} + G_{\gamma\gamma Z_{R}}G_{H_{0}H_{0} Z_{L}},\\
\end{split}\end{equation}
the coupling $G_{\gamma\gamma\gamma, Z}$ and $G_{H_{0}H_{0}\gamma, Z}$ are the triple gauge boson helicity products, $\Gamma_{Z}$ is $Z$ total decay width $\approx. 2.4952\pm 0.0023 GeV$ with two Higgs to polarize gauge boson coupling respectively.
\begin{equation}\begin{split}\label{eq18}
G_{H_{0}H_{0}\gamma} = &- ie\frac{\sqrt{s}}{2}\big{(}2(\vec{p}_{1}\cdot\vec{\epsilon}_{3})(\vec{p}_{1}\cdot\vec{E} - m^{2}_{H}(\vec{E}\cdot\vec{\epsilon}_{3})\big{)},\\
G_{H_{0}H_{0}Z} = &- ie\frac{\sqrt{s}}{4}(3\tan\theta_{w} - \cot\theta_{w})\\
                  &\big{(}2(\vec{p}_{1}\cdot\vec{\epsilon}_{3})(\vec{p}_{1}\cdot\vec{E}) - m^{2}_{H}(\vec{E}\cdot\vec{\epsilon}_{3})\big{)},\\
\end{split}\end{equation}
and
\begin{equation}\begin{split}\label{eq23}
&G_{\gamma\gamma\gamma} =\\
&2e\sin\theta_{w}K_{\gamma\gamma\gamma}\big{(}\frac{s}{2}\big{)}\big{[}(\epsilon_{2}\cdot\epsilon_{3})\big{(}(a-1)\epsilon_{1}\theta k_{1}+2\epsilon_{1}\theta k_{2}\big{)}\big{)}\\
&+ (\epsilon_{1}\cdot\epsilon_{3})\big{(}(a - 1)\epsilon_{2}\theta k_{2} + 2\epsilon_{2}\theta k_{1})\big{)} + (a - 1)(\epsilon_{3}\theta k_{3})(\epsilon_{1}\epsilon_{2})\big{]},\\
&G_{\gamma\gamma Z} =\\
&- 2e\sin\theta_{w}K_{\gamma\gamma Z}\big{(}\frac{s}{2}\big{)}\big{[}(\epsilon_{2}\cdot\epsilon_{3})\big{(}(a-1)\epsilon_{1}\theta k_{1}+2\epsilon_{1}\theta k_{2}\big{)}\big{)},\\
&+ (\epsilon_{1}\cdot\epsilon_{3})\big{(}(a - 1)\epsilon_{2}\theta k_{2} + 2\epsilon_{2}\theta k_{1})\big{)} + (a - 1)(\epsilon_{3}\theta k_{3})(\epsilon_{1}\epsilon_{2})\big{]},\\
\end{split}\end{equation}
where the parameter "a" denotes renormalization constant and the momentum $k^{\mu}_{3}$ = $k^{\mu}_{1}$ + $k^{\mu}_{2}$, with mandelstam variables $s$ = $(k_{1} + k_{2})^{2}$. Renormalization condition requires $a$ = 3. Parity is conserved on the amplitude after exchanging photon polarization $\vec{\epsilon}_{1}$$\to$$\vec{\epsilon}_{2}$ and momentum $\vec{k}_{1}$$\to$$\vec{k}_{2}$ simultaneously.

In Eq.($\hyperref[eq11]{11}$), the measurement of photon energy power spectrum is under the range from the upper bounds to the lower limits $\frac{4m^{2}_{H^{0}}}{E^{2}_{CM}y}$,
\begin{equation}\begin{split}\label{eq24}
&\sigma(\Omega) =\\
&\int^{x_{m}}\int^{x_{m}}_{4m^{2}_{H^{0}}/E^{2}_{CM}y}dxdyf(x)f(y)\bigg{(}\frac{1+\xi(x)\xi(y)}{2}\bigg{)}\sigma.\\
\end{split}\end{equation}
\begin{figure}[htbp] 
   \centering
   \includegraphics[width=3.2in]{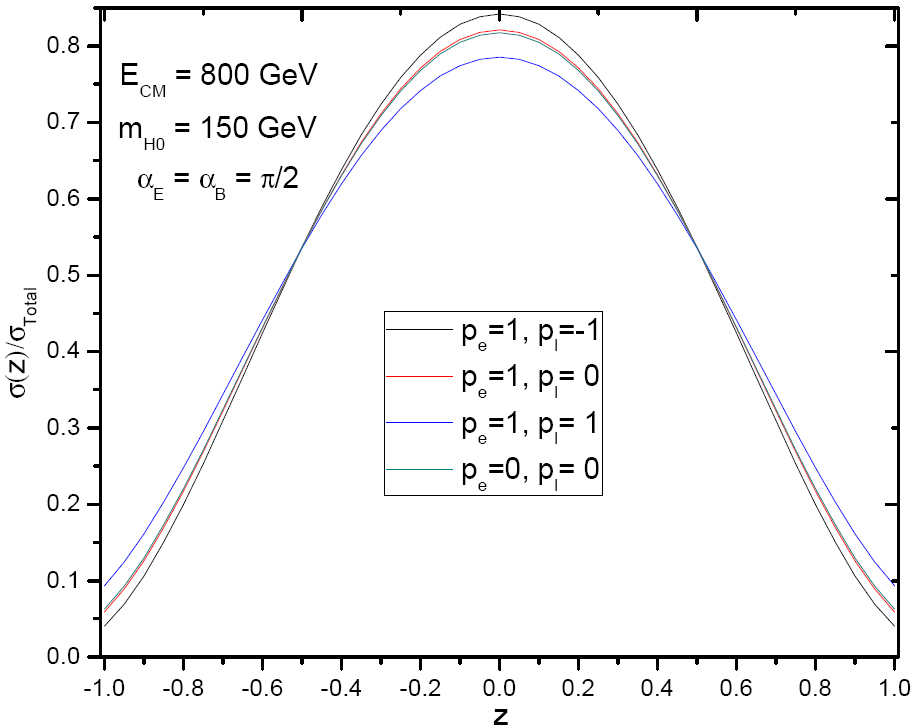}
   \includegraphics[width=3.2in]{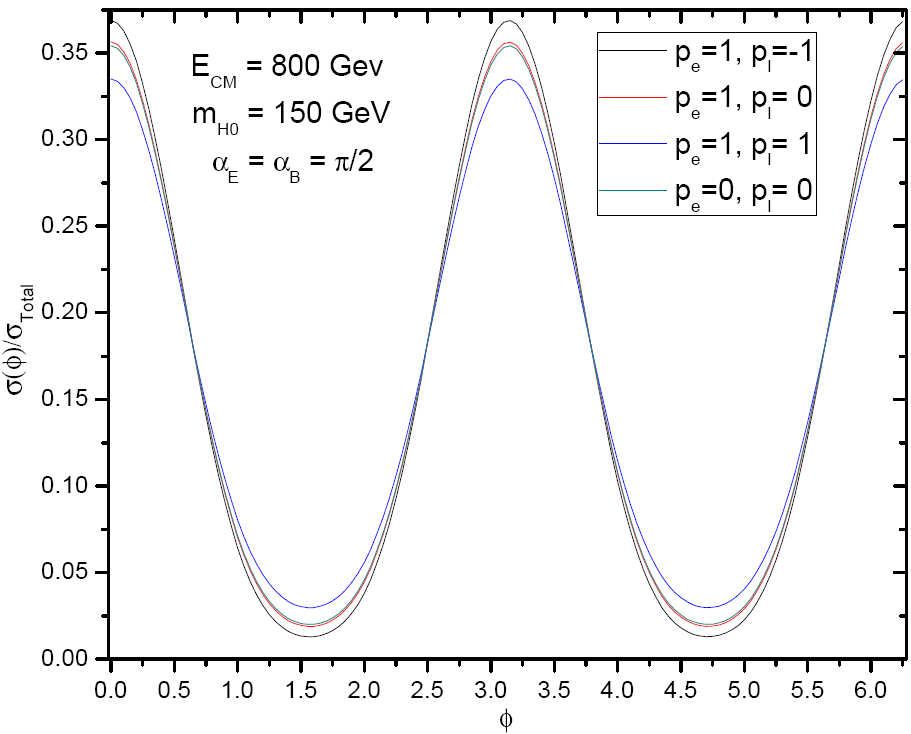}
   \caption{The differential cross section on the energy level $\Lambda_{c}$ = 1 TeV, and the central energy $E_{CM}$ = 800 GeV with Higgs mass $m_{H^{0}}$ = 150 $GeV$, the distribution is symmetric at the point $z$ = 0 and at $\phi$ = 0.}
   \label{fig:gh0dp}
\end{figure}
In Fig.(\hyperref[fig:gh0dp]{4}), the ratio of polarized and unpolaried cross section are sensitive to be changed on difference polarized laser beam. The transverse and linear incident laser photon both are generated on the photon collider technique~\cite{tesla}. In the common standard model analysis, the $P$ even particle couples to a linear polarized photon with the maximal strength of parallel polarization vector. The $P$ odd particle is perpendicular to the photon polarization. The neutral Higgs particles, $H^{0}$, and $h^{0}$ are $0^{++}$ quantum number without fermion fields, and the $0^{++}$ and $0^{--}$ states in the model taking fermion field into account. The existence of $P$ odd and $C$ odd states result from fermion sector violates parity and charge conservation on one loop chiral triangle diagrams, e.g. anomaly magnetic moment and electric dipole moment.
\begin{figure}[htbp] 
   \centering
   \includegraphics[width=3.2in]{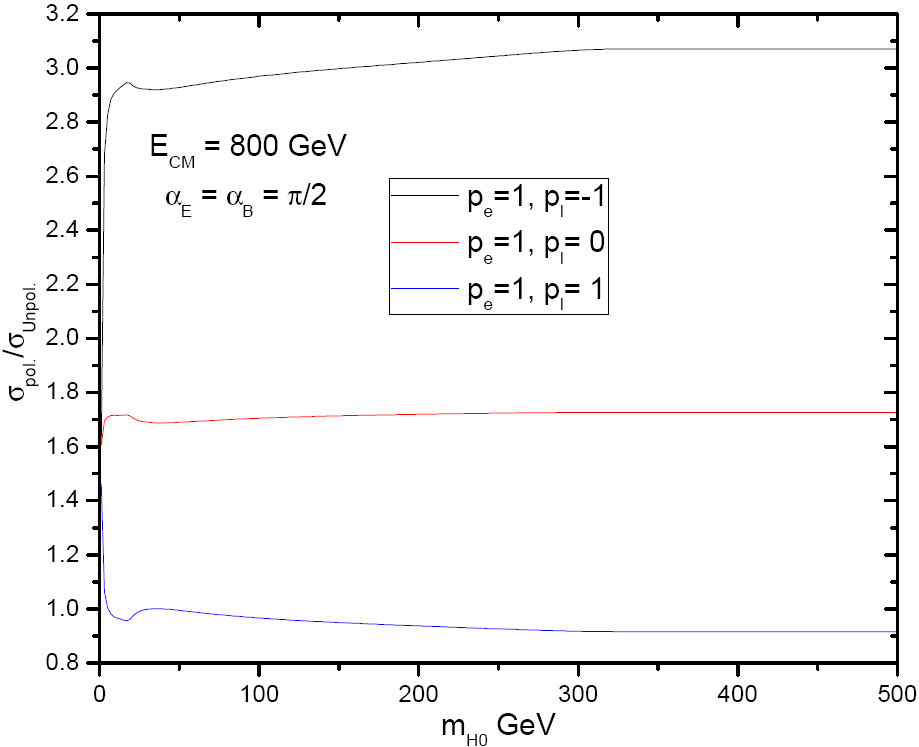}
   \caption{The total cross section with the direction of background magnetic field, at the point of $\alpha$ = $\frac{\pi}{2}$ under the zeroth total angular moment process without violating parity. The maximum strength distribution is under the pole $p_{e}$ = 1 and $p_{l}$ = -1.}
   \label{fig:gh0mh}
\end{figure}

On the $\theta$ deformed spacetime, the background magnetic field couples to polarized photon intrinsic properties. We extract the background electric and magnetic products, $\epsilon^{1, 2}_{\mu}\theta^{\mu\nu}k^{3}_{\nu}$, $\epsilon^{1}_{\mu}\theta^{\mu\nu}k^{2}_{\nu}$,
and $\epsilon^{2}_{\mu}\theta^{\mu\nu}k^{1}_{\nu}$, from the couplings, Eq.($\hyperref[eq23]{23}$), and Eq.($\hyperref[eq24]{24}$), and the momentum couples to background field from, $\vec{p}_{1,2}\cdot\vec{E}$, terms. The volume of $\vec{B}\cdot(k_{i}\times\epsilon_{i,j})$ production is such as the momentum and circle polarization rotated objects around background $\vec{B}$ field, the electric field is similar devoted on altering momentum translational invariance.

In Fig.($\ref{fig:gh0mh}$), the range for neutral Higgs particle, $H^{0}$, the mass spectrum on the maximum mode, $p_{e}$ = 1 and $p_{l}$ = -1, generates a maximum analysis results. The integration is restricted in getting a positive final particle velocity $v$. Integrating out the full x-beam channel and y-beam channel energy range\footnote{The minimal and maximum range are $\sqrt{1 - 4m^{2}_{H^{0}}/E^{2}_{CM}xy}$ and $x_{m}$ respectively.}, the polarized mode and unpolarized consequence is manifestly presented. In the two sides incoming laser site, if the field direction is paralleled to the z-axis, the distribution is minimum contributed. Under theoretical prediction, the massless particle cannot correct power distribution on its mass spectrum. The total devotion attributes to the polarized photon gas. The coupling in Fig.($\hyperref[FR]{1}$) is $CP$ conserved, but, violates $C$ and $P$ respectively by including fermion field. Without fermion field, $C$ and $P$ can be violated or conserved respectively in $Z$ gauge particle (see Tab.II).

We focus on $\alpha_{B}$ = $\alpha_{E}$ = $\frac{\pi}{2}$ and set $\beta_{B}$ = $\beta_{E}$ = 0, even though the dramatic field is mainly to influence the process, but no parity effect is generated under the field interaction with photon polarization. Intuitively, if no $P$ violation effect on the first order $\theta$ deformed term, each helicity state is essentially dedicated to the process.

\begin{figure}[htbp] 
   \centering
   \includegraphics[width=3.2in]{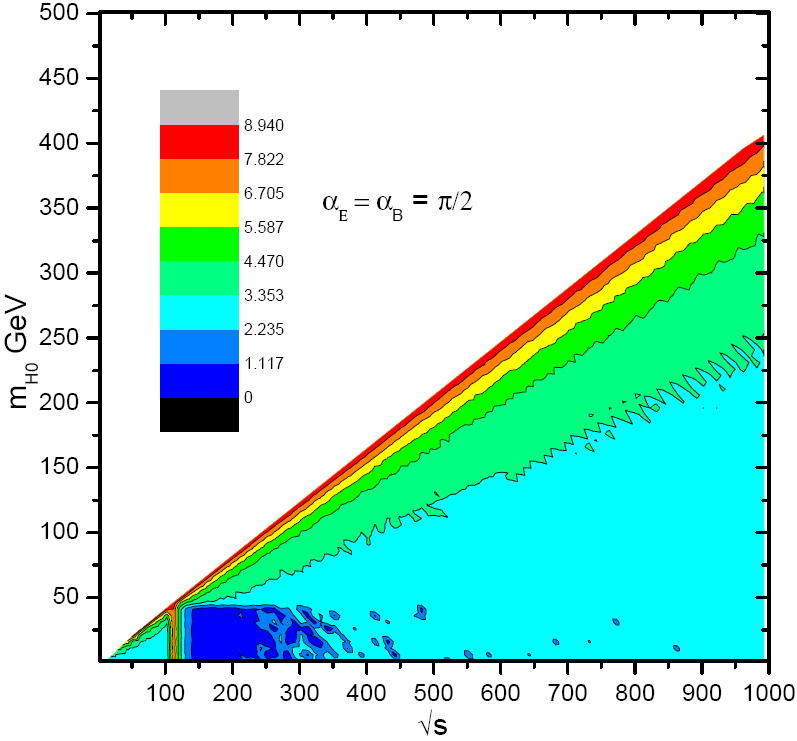}
   \caption{The fractional power rate at the pole, $p_{e}$ = 1 and $p_{l}$= - 1, is sensitive changed under the central energy range 100 GeV $\sim$ 500 GeV. Higgs particle mass can be lower on 50 GeV in Tab.III. The barrier wall around 100 GeV separates the two range between lower Higgs mass production, $\sim$50 GeV, and higher Higgs mass range, $>$50 GeV. As the $Z_{0}$ particle mass is around 0.1 TeV, the collapsed power spectrum is dominated on massive gauge boson resonance. The result is very minuscule to change the direction of the background electric field, whether with/without changing the direction of the background magnetic field.}
   \label{fig:gh0cemh}
\end{figure}

The total strength is a function as to the direction of background field $\alpha_{B}$ and $\alpha_{E}$, between observer and background field direction. In Fig.(\hyperref[fig:gh0mh]{5}), the total cross section is sensitive influenced by photon and electron beam. Furthermore, the incoming electron polarization and laser photon beams are mainly controlled on pure laser photon. In the case, $p_{e}$ = 1 and $p_{l}$ = -1, its maximum power distribution is around the incident point (IP). The electron and incoming laser photon are the same angular momentum eigenstates. In comparing to the minimum case, $p_{e}$ = 0 and $p_{l}$ = 0, the ratio of polarized and unpolarized beams, Fig.($\hyperref[fig:gh0cemh]{6}$), the power distribution is minimum in the central energy between 100 GeV$\sim$ 500 GeV. 

Under low incoming energy range, the collapse effects is generated from the accelerated phenomena under power law conservation condition. The low energy range photon gauge boson can decomposed into two spin-$\frac{1}{2}$ particles in the interaction of background $\theta_{\mu\nu}$ patch twisted spacetime. Some articles~\cite{Belotsky:2002ym, Godfrey:2004gv} consider a fourth generation fermion doublet enhance to reach a mixing area for searching lost neutrino mass. In which, it is a complete phenomenon of degenerated energy loss form spinor decouple. However, its chiral properties is broken on this produced decouple effects. Hence, the degenerated energy power spectrum redefines a mixed spin-$0$ chiral violated resonance. There numerous light Higgs particles are generated above the 0.1 GeV in Fig.(6). The double neutral Higgs production is a regular spin-0 particle without mixing $P$ or $C$ asymmetry effects. The fractional power distribution devotes on the created light neutral Higgs particles is manifestly. In this bounds, the central energy larger than 2$m_{H^{0}}$, the required power is maximum devoted on creating $H^{0}$ particles.

\subsection{$\gamma\gamma\to H^{+}H^{-}$}
In this and next section, we compare $\gamma\gamma\to H^{+}H^{-}$ and $H^{0}A^{0}$ with $H^{0}H^{0}$ process. The 2HDM~\cite{Gunion:1984yn, Dawson:1989ws, Ellis:1988er, Muhlleitner:2001zv} considers two complex scalar field on the different vacuum $v_{1}$ and $v_{2}$ in the $\xi$ phase connects with each other. On non-supersymmetric model, the expression of doublet scalars on these vacuums are all down type form,
\begin{equation}\label{eq24}
\phi_{1} = \begin{pmatrix} 0 \\ v_{1}\end{pmatrix},\quad\phi_{2} = \begin{pmatrix}0 \\ v_{2}e^{i\xi}\end{pmatrix},
\end{equation}
in considering the general potential whether $CP$ is conserved or not. Rotating the field into neutral $H^{0}$ Higgs and its orthogonal state, the goldstone particle $G^{0}$, and the standard massive Higgs particle are generated on the model building.

Charged scalar Higgs, $H^{+, -}$, and charged goldstone field ,$G^{+, -}$, and other neutral fields $A^{0}$ and $h^{0}$, are produced from the imaginary part and its real part. The down type doublet scale are also taking into account. There are $CP$ even particles, $H^{+, -}$, $H^{0}$, and $h^{0}$, and $CP$ odd particle $A^{0}$ on this model. The mass term of the neutral Higgs and the $CP$ even chargeless scalar $h^{0}$ are in order to diagonalize the mass mixing matrix with the rotation angle $\alpha$ under the ground state fluctuations $Re\phi_{1}$ and $Re\phi_{2}$. The only associated relation between two vacuum $v_{1}$ and $v_{2}$ is $\tan\beta$ = $\frac{v_{2}}{v_{1}}$.

We set the $\cos(\beta-\alpha)\ll\sin(\beta-\alpha)$, the diagram of $A^{0}h^{0}$ case is proportional to the $A^{0}H^{0}$ by the factor $\tan^{2}(\beta-\alpha)$. The data in $Particle~Data~Group$ shows that the ratio $\tan\beta$ in ABBOTT 99E\footnote{ABBOTT 99E search for a charged Higgs boson in top decays in p$\bar{p}$ collisions at $E_{CM}$= 1.8 TeV, by comparing the observed t$\bar{t}$ cross section and assuming the dominant decay $t\to bW^{+}$ with theoretical expectation.} is separated by the two range, $\tan\beta\lesssim~1$,$50<m_{H^{+}}(GeV)\lesssim~120$, and $\tan\beta\gtrsim$ 50, as 50$< m_{H^{+}}(GeV)\lesssim 160$. In ABAZOOV 02B\footnote{ABAZOV 02B search for a charged Higgs boson in top decays with $H^{+}\to\tau^{+}\nu$ at central energy , $E_{CM}$, = 1.8 TeV}. For $m_{H^{+}}$ = 75 GeV, the range $\tan\beta>32.0$ is excluded at 90$\%$CL. For $\tan\beta$ values above 140, the excluded mass range extends to over 140 GeV scale. In AFFOLDER 001 \footnote{Searching a charged Higgs boson in top decays with the $H^{+}\to\tau^{+}\nu$ in hadron collider $p\bar{p}$ at $E_{CM}$ = 1.8 TeV.}, the excluded mass range extends to over 120 GeV for $\tan\beta$ values above 100 and $B(\tau\nu)$ = 1. By assuming the mixing angle $\alpha$ is small, the dominated area towards into search the fractional rate of $\tan\beta$ (see Tab.III).

\begin{figure}[htbp] 
   \centering
   \includegraphics[width=2in]{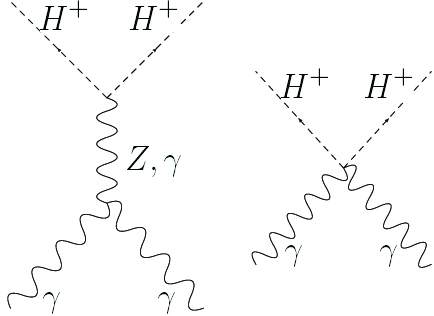}
   \caption{$\gamma\gamma\to Z, \gamma\to H^{+}H^{-}$.}
   \label{fig:gh+h-}
\end{figure}

We assume charged Higgs mass $m_{H^{\pm}}$ = 150 GeV at $E_{CM}$ = 800 GeV. The diagram in Fig.($\ref{fig:gh+h-}$) contains middle $Z$ and $\gamma$ resonance. Neutral Higgs production on noncommutative model is derived from Fig.($\ref{fig:FR}$). In loop calculation, the effective lagrangian of standard model predicted in $\phi^{0}\to\gamma\gamma$ couplings\footnote{The loop diagrams contain chargeless particle and charged particle.
\begin{equation*}\label{eq25}
\mathcal{L}_{int} = -\frac{gm_{f}}{2m_{W}}\bar{\psi}\psi\phi^{0} + gm_{W}W^{+}_{\mu}W^{\mu-}\phi^{0} - \frac{gm^{2}_{H}}{m_{W}}H^{+}H^{-}\phi^{0} + \cdots
\end{equation*}
The devotion for spin- 1, spin-0, and spin-$\frac{1}{2}$ middle states that bosonic symmetry is violated on the one loop level.} is sensitively dependent on Higgs mass. This program is a standard process working on noncommutative triple gauge boson coupling $\gamma\gamma Z^{0}$ and $\gamma\gamma\gamma$, and standard model, $Z^{0} H^{+}H^{-}$, predicted one. The couplings between photon helicity polarization to background field, $H^{+}H^{-}$ pair obtains a minimum in comparing to $H^{0}H^{0}$ and $A^{0}H^{0}$ processes.

The amplitude on the gamma ray helicity state is contributed by first order $\theta$ deformed term. It appears that the total cross section is single two order deformation. The order magnitude around the electroweak scale is approaching to $\frac{s}{\Lambda^{2}_{C}}$ extension.
\begin{equation}\begin{split}\label{eq26}
&M_{\gamma\gamma\to H^{+}H^{-}} =\\
&2\bigg{(}\frac{k_{\gamma\gamma Z}\cos 2\theta_{W}}{s-m^{2}_{Z}-im_{Z}\Gamma_{Z}} - \frac{K_{\gamma\gamma\gamma}\sin 2\theta_{W}}{s}\bigg{)}\\
&\bigg{(}\square_{L}[p_{2}\cdot\epsilon_{3R}-p_{1}\cdot\epsilon_{3R}] + \square_{R}[p_{2}\cdot\epsilon_{3L}-p_{1}\cdot\epsilon_{3L}]\\
&\qquad\qquad\qquad\qquad\qquad\qquad\qquad\qquad\qquad + 2i \epsilon_{1}\cdot\epsilon_{2}\bigg{)},\\
\end{split}\end{equation}
in which $\square_{L}$ and $\square_{R}$ are
\begin{equation}\begin{split}\label{eq27}
\square_{i} &=(a-1)\bigg{(}\frac{s}{2}\bigg{)}(\epsilon_{1}\cdot\epsilon_{2})(\epsilon_{3i}\theta k_{3})\\
                     &+ \bigg{(}\frac{s}{2}\bigg{)}\big{(}(a-1)\epsilon_{1}\theta k_{1} - 2\epsilon_{1}\theta k_{2}\big{)}\epsilon_{2}\cdot\epsilon_{3i}\\
                     &+ \bigg{(}\frac{s}{2}\bigg{)}\big{(}(a-1)\epsilon_{2}\theta k_{2} - 2\epsilon_{2}\theta k_{1}\big{)}\epsilon_{1}\cdot\epsilon_{3i},\\
\end{split}\end{equation}
the index $i$ denotes right-handed and left-handed circle polarization. In order to avoid considering the change of the interaction between background field and the mediate gauge bosons. We contract the gauge boson polarization with the coupling tensor immediately. Extracting separated terms, Eq.($\hyperref[eq22]{22}$) shows that the coupling between particle momentum and gauge boson polarization on the content are important to decide the physical results. The integration of energy fractional rate from upper to lower limits are
\begin{equation}\begin{split}\label{eq23}
&\sigma(\Omega) =\\
&\int^{x_{m}}_{0}\int^{x_{m}}_{4m^{2}_{H^{\pm}}/E^{2}_{CM}y}dxdy f(x)f(y)\bigg{(}\frac{1 + \xi(x)\xi(y)}{2}\bigg{)}\sigma,\\
\end{split}\end{equation}
where the range is up to $x_{m}$ threshold ratio and down to the particle creation condition, the velocity has to be $\ge$ 0 at least. The $Z$ resonance cannot take a constraint on energy range from $m_{Z}$ to TeV scale. The photon power spectrum is a function restricted on the range, the lower bound is under the causality constraint and upper limits is lower the machine maximum power restrictions.

In the photon photon backreact collider technology, the total cross section associates with the incoming laser energy rate, and constrained by the physical quantity. The parameter in this process is the direction of background field, $\vec{B}$ and $\vec{E}$, and the scalar mass under the assumption of the perturbation scale, $\Lambda_{C}$, is around 1 TeV. We set $\beta_{B}$ = $\beta_{E}$ = 0, and the central energy, $E_{CM}$, at 800 GeV with the total cross section depends on $\alpha_{B}$ and $\alpha_{E}$ parameters. It is an abnormal imposed variables, due to background field influence.

\begin{figure}[htbp] 
   \centering
   \includegraphics[width=3.2in]{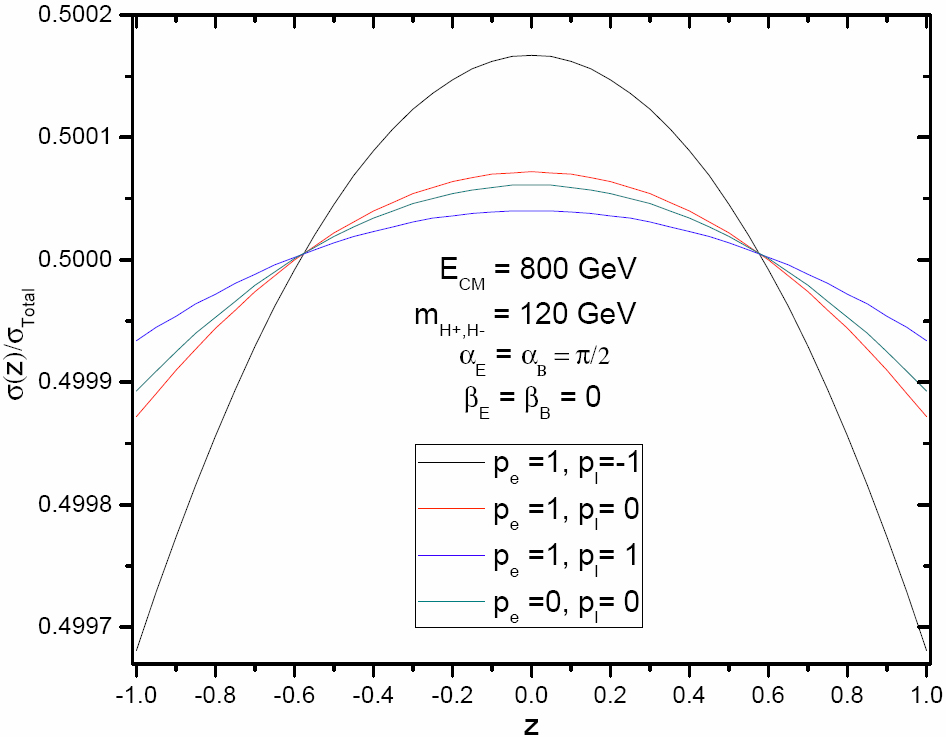}
   \includegraphics[width=3.2in]{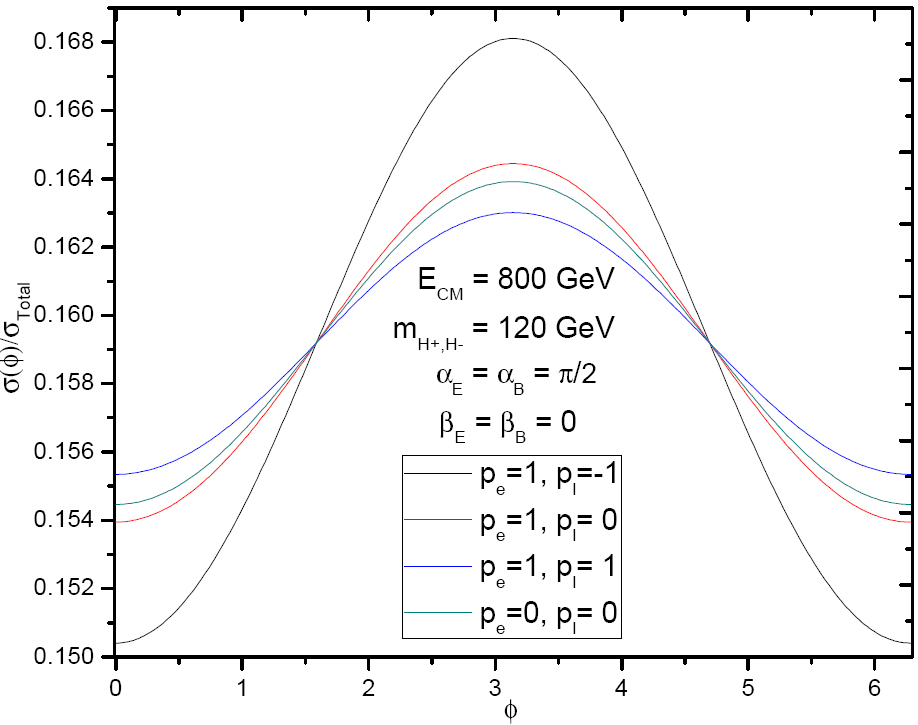}
   \caption{The differential cross section on the energy level $\Lambda_{c}$ = 1TeV, and the central energy $E_{CM}$ = 800 GeV with Higgs mass $m_{H^{\pm}}$ = 120 GeV. The distribution on the point, z = 0 and $\phi$ = 0, the probability is maximum distributed.}
   \label{fig:ghd}
\end{figure}

The global angle, $\Omega$, is similar as a variables of the probability of charged Higgs boson creating process. The incoming photon and electron beam, the polarization sensitively controls the experiments consequence. The maximum polarized value in Fig.($\hyperref[fig:ghd]{8}$) is around 0.5018 in first figure and 0.177 in second figure. The background field plays a ridicule role on the total cross section. The global consequence, the angle $\alpha_{E}$ and $\alpha_{B}$, and $\beta_{E}$ and $\beta_{B}$ are existed on the experiment parameters. They are independent with the detector direction. Naturally, the distribution cannot be shifted by moving to different sites. The pole $p_{e}$ = 1, $p_{l}$ = -1 is a maximum contribution, the pole $p_{e}$ = 1, $p_{l}$ = 1 produce a minimum devotion on the final results. We set $\alpha_{E}$ = $\alpha_{B}$ = $\frac{\pi}{2}$, and $\beta_{E}$ = $\beta_{B}$ = 0 on the $\gamma~Z$ channel.

\begin{figure}[htbp] 
   \centering
   \includegraphics[width=3.2in]{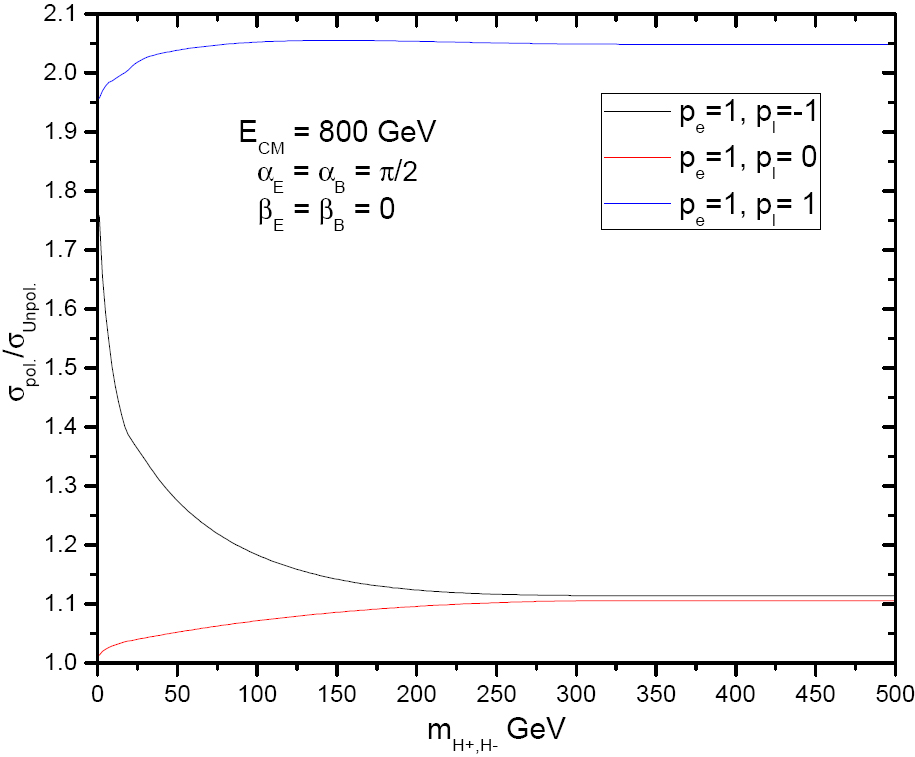}
   \caption{The total cross section with the direction of background magnetic field, on the point of $\alpha$ = $\frac{\pi}{2}$ under the zero total angular moment process, there are no parity violation evidence, the maximum strength distribution is under the $p_{e}$ = 1 and $p_{l}$ = -1 beam. The polarized polar under $p_{l}$ = -1 and $p_{l}$ = 1 are over crossed nearby the point $m_{H^{+}, H^{-}}\sim$ 0. The fractional rate of the un-polarized photon laser beam cannot be polarized at beginning.}
   \label{fig:ghmh}
\end{figure}

The mass spectrum in Fig.($\hyperref[fig:ghmh]{9}$) is very different as taking $m_{H^{\pm}}$ =0. It is result from the high frequency photon remnants. The modified energy is stored into background field interaction with photon polarization. The $p_{e}$ =0, $p_{l}$ = 0 pole is minimal contributed on the polarized rate. Contrastively, the pole $p_{e}$ = 1 and $p_{l}$ = 1 is maximum devoted at $E_{CM}$ = 800 GeV. It is a very dramatic phenomenon in probing charged Higgs at high energy gamma ray collision. In the limit range, photon polarization $p_{l}$ = 0 and -1 is approaching to the same rate. The maximum power fractional devotion is posed at the pole $p_{1}$ = 1.

\begin{figure}[htbp] 
   \centering
   \includegraphics[width=3.2in]{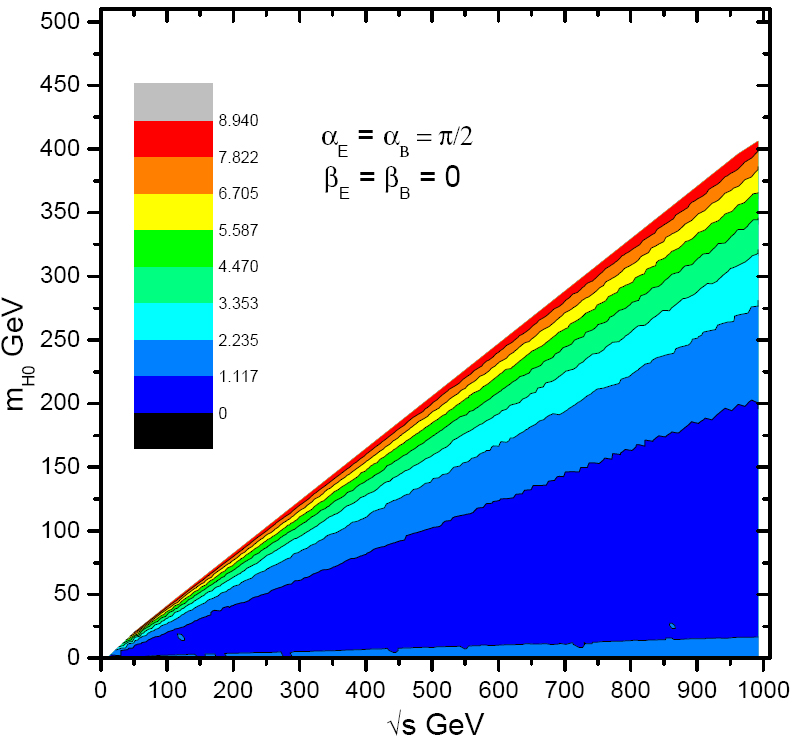}
   \caption{The fractional power distribution is posed on $\alpha_{E, B}$ = $\frac{\pi}{2}$ and $\beta_{E, B}$ = 0. The regular devotion is working on the restriction of $E_{CM}$ = $m_{H^{\pm}}$. Near the deadline, it is maximum devoted.}
   \label{fig:ghcemh}
\end{figure}

In Fig.($\hyperref[fig:ghcemh]{10}$), it is a regular form of the process. The minimum is under the right-down corn, and the maximum is in the high slope area. Under the minimum momentum final state, particle power is maximum distributed. At the lower mass spectrum, particle is vastly numerous produced in the fine changes on the central energy. Along the defined background direction, it does not vary dramatically. At low energy scale to high energy scale, power spectrum is similar as the same linear function from the massless point to 400 GeV mass scale. Linear distribution in power spectrum implies that the relation between incoming high energy photon frequency is uniformed devoted on the particle angular distribution. The fractional rate in photon gas power spectrum denotes that particle mass is a smooth function with its angular distribution.

\subsection{$\gamma\gamma\to H^{0}A^{0}$}
In comparing neutral Higgs and charged Higgs production process, in this section we introduce a $CPV$ scalar particle $A^{0}$ in the $\gamma\gamma$ collider background. The general model building, scalar potential in 2DHM is $CP$ odd as to consider the imagine part of the factor $\sin\xi$. It is a $CP$ even by ignoring the factor from the process. Splitting the complex field $\phi_{1}$ and $\phi_{2}$ into real part and imagine part, there two extra degrees of freedom are indicated into particle creation. The $CPV$ particle $A^{0}$ induces to the imagine field, and the real part produces a $CP$ even particle $h^{0}$.

In the high energy gamma ray background creation, it contains a fluently pure high energy gamma ray simulation analysis. In Tab.II, the quantum number $A^{0}$ particle is under $0^{+-}$ state in the model without lepton and quark sector. In which $CP$ even and $CP$ odd melted vacuum, the $CPV$ effect is embedded into a non-equalibrium state. The $P$ violated coupling $ZH^{0}A^{0}$ is not a physical quality due to unitarity and bosonic symmetry is inherent property in vacuum. Bosonic condition considers $C$ and $P$ or $CP$ conservation. $CPT$ preserves on the thermal equilibrium, and the conservation law imposes on $\mathcal{L}(t)$ is the same as exchanging $t\to -t$.  $P$ conservation is a natural effects in $\gamma\gamma$ to $H^{0}A^{0}$ process.

In discussing the background in high frequency photon gas, the expectative influence is $P$ violated and $C$ violated phenomenon. On this process $C$ violation is visible, but, $P$ is disappeared, since, no first order $\theta$ on the cross section with no helicity state of final created fields. The $A^{0}$ mass distribution on a range of its the mass spectrum depends on
background field direction, expansion scale, and its incoming energy. In supersymmetry model, the extra $H^{0}_{1}$ and $H^{0}_{2}$ scalar particles are added into model in considering an up type and a down type doublet. The only unknown variables are $m_{A^{0}}$ and $\tan\beta$ in 2DHM.

\begin{figure}[htbp] 
   \centering
   \includegraphics[width=1in]{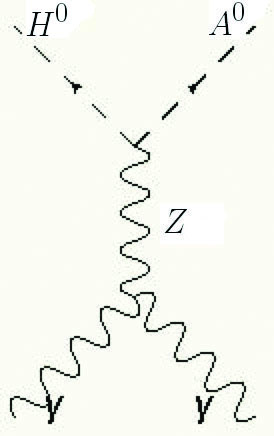}
   \caption{$\gamma\gamma\to H^{0}A^{0}$.}
   \label{fig:gha}
\end{figure}

The momentum of final state particles $H^{0}$ and $A^{0}$ on $CM$ frame are $|\vec{p}_{H^{0}}|$ = $|\vec{p}_{A^{0}}|$ = $\frac{\sqrt{s}}{2}|\vec{v}|$ with the velocity $|\vec{v}|$ is
\begin{equation}\label{eq29}
v = \sqrt{1+\bigg{(}\frac{m^{2}_{H^{0}}}{s} - \frac{m^{2}_{A^{0}}}{s}\bigg{)}^{2} - 2\bigg{(}\frac{m^{2}_{H^{0}}}{s} + \frac{m^{2}_{A^{0}}}{s}\bigg{)}},
\end{equation}
the lower bound is onto the mathematics condition. Its amplitude is
\begin{equation}\begin{split}\label{eq30}
&M_{\gamma\gamma\to H^{0}A^{0}} =\\
&-2i\frac{K_{\gamma\gamma Z}}{s - m^{2}_{Z} - im_{Z}\Gamma_{Z}}\\
&\bigg{(}\square_{L}[p_{2}\cdot\epsilon_{3R}-p_{1}\cdot\epsilon_{3R}] + \square_{R}[p_{2}\cdot\epsilon_{3L}-p_{1}\cdot\epsilon_{3L}]\bigg{)},\\
\end{split}\end{equation}
where $\square_{i}$ is introduced in Eq.($\hyperref[eq28]{28}$), and the $\Gamma_{Z}$ is the total $Z$ decay width. The triple gauge boson coupling, $K_{\gamma\gamma Z}$, sets to - 0.3 on the restriction of model building. This process is fully contributed by noncommutative spacetime geometry, under $\theta$ second order amplitude,
\begin{equation}\begin{split}\label{eq31}
&\sigma(\Omega) =\\
&\int^{xm}\int^{xm}_{(m_{H^{0}}+m_{A^{0}})^{2}/E^{2}_{CM}y}dxdyf(x)f(y)\bigg{(}\frac{1+\xi(x)\xi(y)}{2}\bigg{)}\sigma.\
\end{split}\end{equation}
\begin{figure}[htbp] 
   \centering
   \includegraphics[width=3.2in]{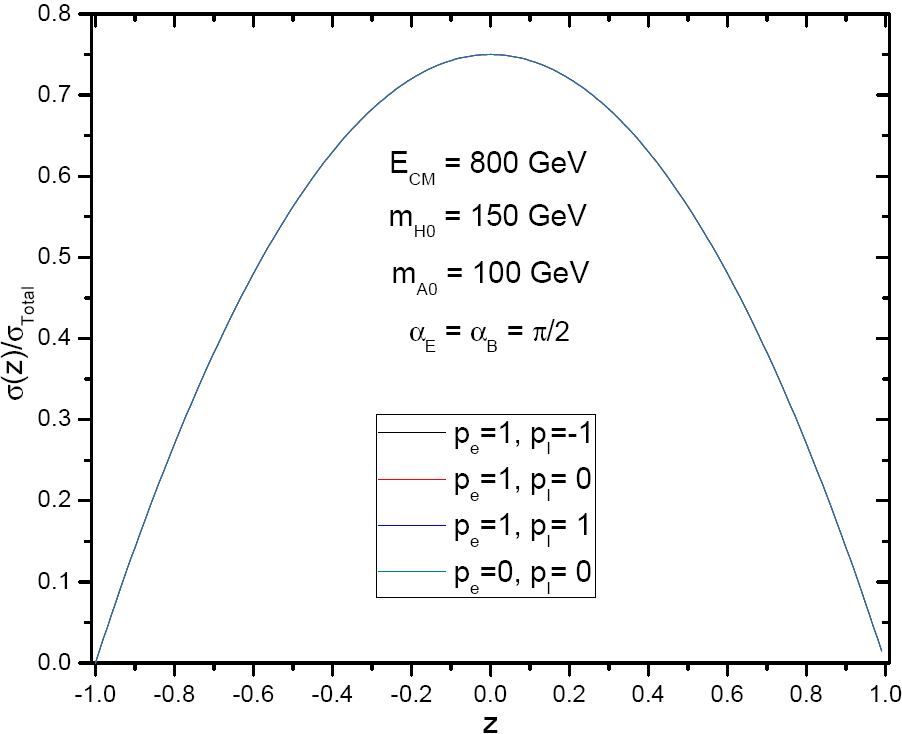}
   \includegraphics[width=3.2in]{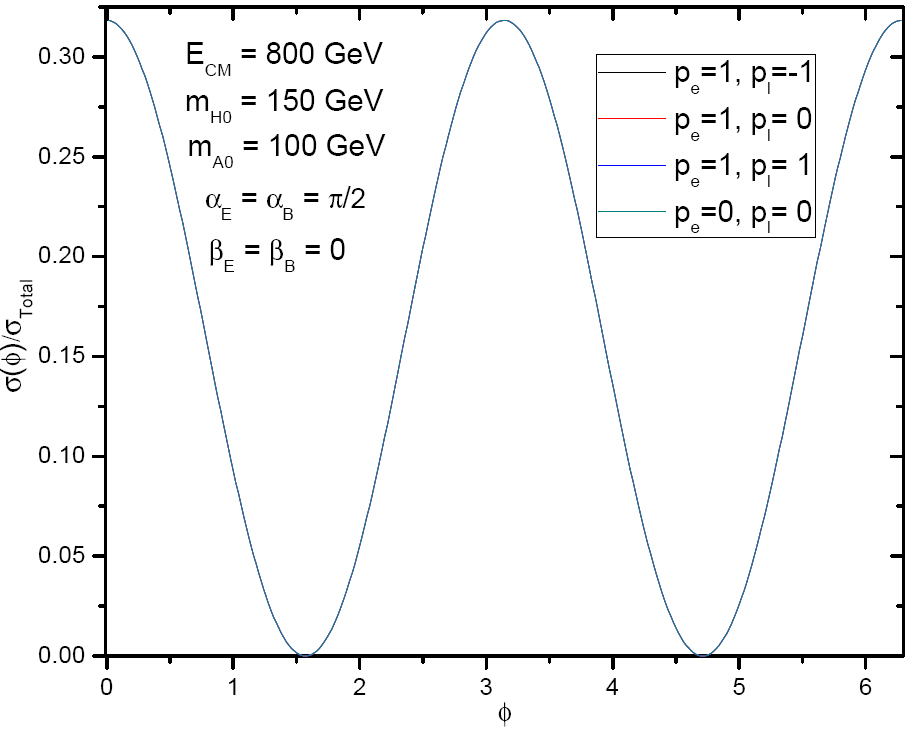}
   \caption{The differential cross section, $\sigma(\phi)$, is a functional of $\sin\alpha_{E,B}$ and $\cos(\beta_{E} - \phi)\&\cos(\beta_{B} - \phi)$, and $\sigma(z)$ depends on $\sin\alpha_{E, B}$. The perturbative scale $\Lambda_{C}$ = 1 TeV, and the central energy $E_{CM}$ = 800 GeV with Higgs mass $m_{H^{0}}$ = 150 GeV and  $m_{A^{0}}$ = 100 GeV by imposing the background parameter $\alpha_{E, B}$ = $\frac{\pi}{2}$. The contribution of all pole incoming high energy gamma ray are the same as each others.}
   \label{fig:ghad}
\end{figure}

The symmetry distribution, Fig.($\hyperref[fig:ghad]{12}$), manifestly presents the dominant distribution incident on the perpendicular to the incoming electron laser beam, the result is independent on the background field direction $\alpha_{E}$ or $\alpha_{B}$. The polarized rate is maximum at 7.5. The sinusoid function associates with the background direction $\beta_{E}$ and $\beta_{B}$ angle. We set $\beta_{E}$ = $\beta_{B}$ = 0 for convenient. The polarized rate is the same for all incoming laser pole, the maximum rate is upper 3.2 for $m_{A^{0}}$ = 100, and $m_{H^{0}}$ = 150

\begin{figure}[htbp] 
   \centering
   \includegraphics[width=3.2in]{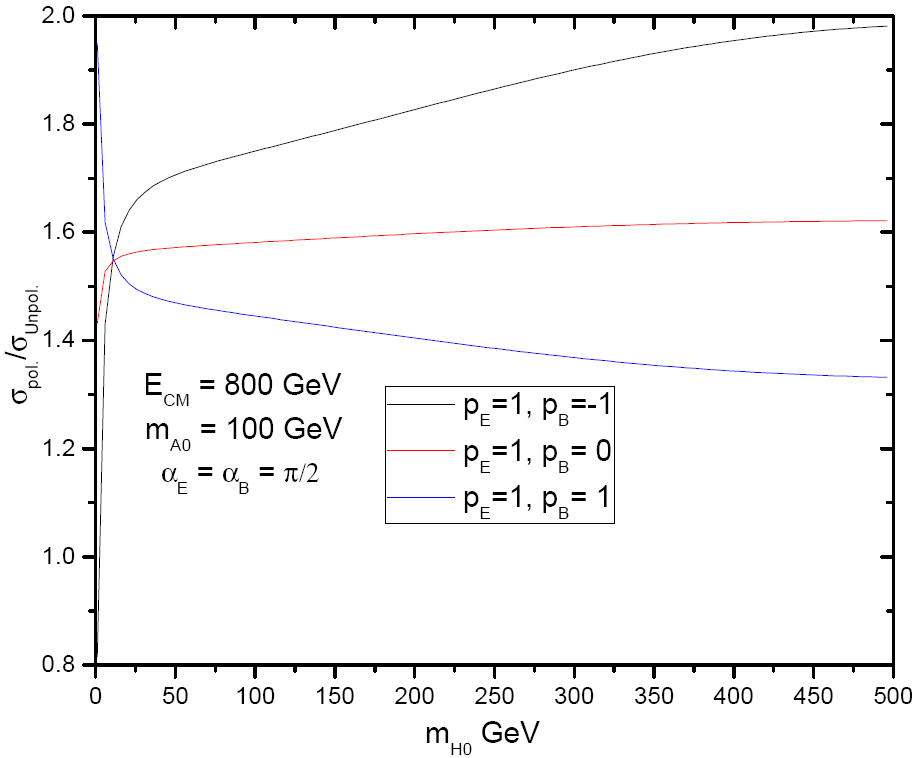}
   \includegraphics[width=3.2in]{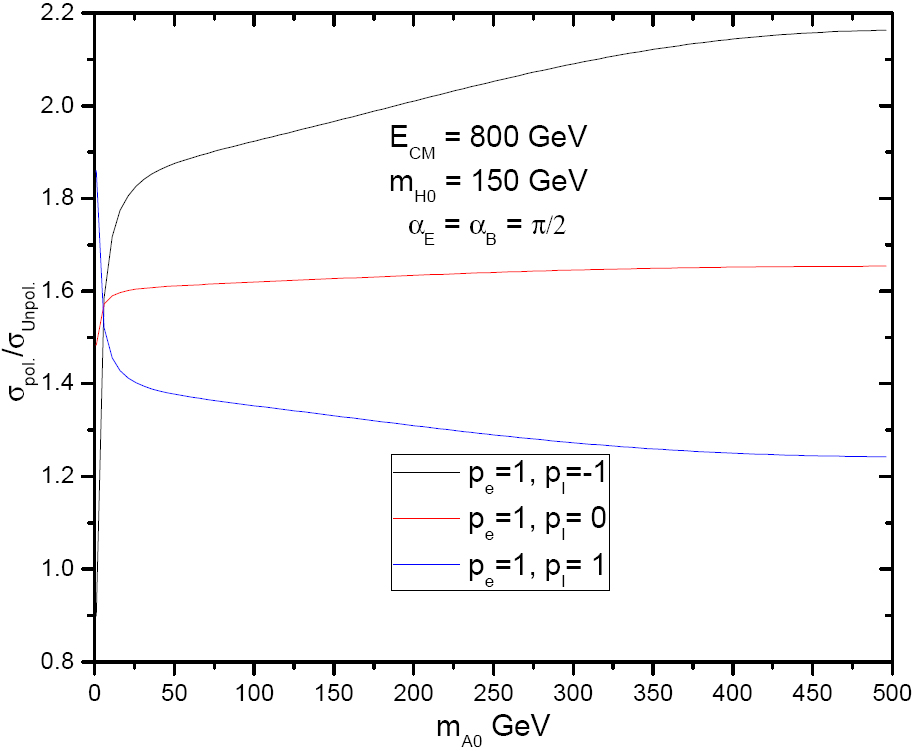}
   \caption{The fractional power rate with the background magnetic field and electric field direction on the angle $\alpha_{E}$ = $\alpha_{B}$ = $\frac{\pi}{2}$, the maximum contribution on the mode $p_{e}$ = 1 , $p_{l}$ = -1, implies that all of the charged scalar particles in 2DHM are as same as each other as they are massless.}
   \label{fig:ghama}
\end{figure}

The $A^{0}$ and $H^{0}$ mass spectrum are crossed around the 80 GeV, in Fig.($\hyperref[fig:ghamh]{13}$). At the first pole, it contributes a maximum energy momentum power in backreact $Compton$ like scattering. It shows that increasing the particle mass the fractional rate is sinuously changed. The polarized fractional rate is the same as each others as to the particle mass $m^{H^{0}}$ and $m_{A^{0}}$.  At the mass scale up to 500 GeV level, the maximum fractional value is around 2.2 and the minimum is down to 1.24. Formally, the incident energy at several
TeV and the perturbative scale assumes at 10 TeV the final luminosity is minuscule by the proportional incoming energy with scale $\Lambda_{C}$. we set the $E_{CM}$ is 800 GeV and $\Lambda_{C}$ at 1TeV level. The incident phenomenon is vivid and observed.

The central energy $E_{CM}$ and Hggs mass spectrum, Fig.($\hyperref[ghasa]{13}$) shows that the relation between central energy, $m_{H^{0}}$ and $m_{A^{0}}$ are similar as the power range distribution at larger energy scale. It is different at the low scale power spectrum. At low photon frequency, Higgs particle, $H^{0}$, expresses a limit distribution on the range lower than 100 GeV. The angular power devotion by $m_{A^{0}}$ presents a larger density at the central energy lower than 500 GeV. The sensitive phenomenon in the photon power distribution on $\gamma\gamma\to H^{0}A^{0}$ process depends on incoming high energy gamma ray. On the high frequency power range, the $\gamma\gamma$ collider presents a uniformed contribution on the polarized and unpolarized events. In controlling the background field direction, there are some modifications are presented, due to the total cross section depends on $\sin^{2}\alpha_{E, B}$ and $\cos^{2}\alpha_{E,B}$ terms.

\begin{figure}[htbp] 
   \centering
   \includegraphics[width=3.2in]{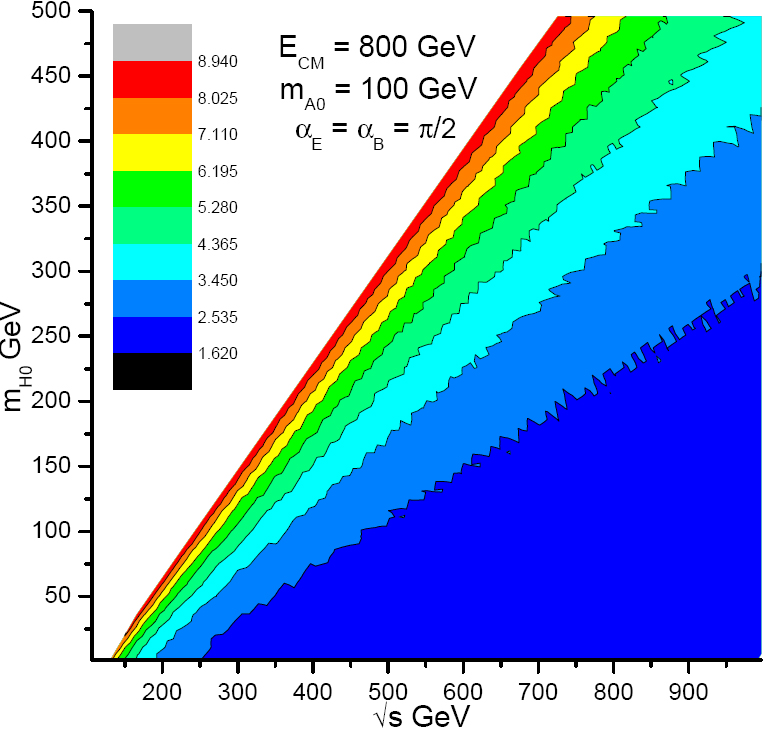}
   \includegraphics[width=3.2in]{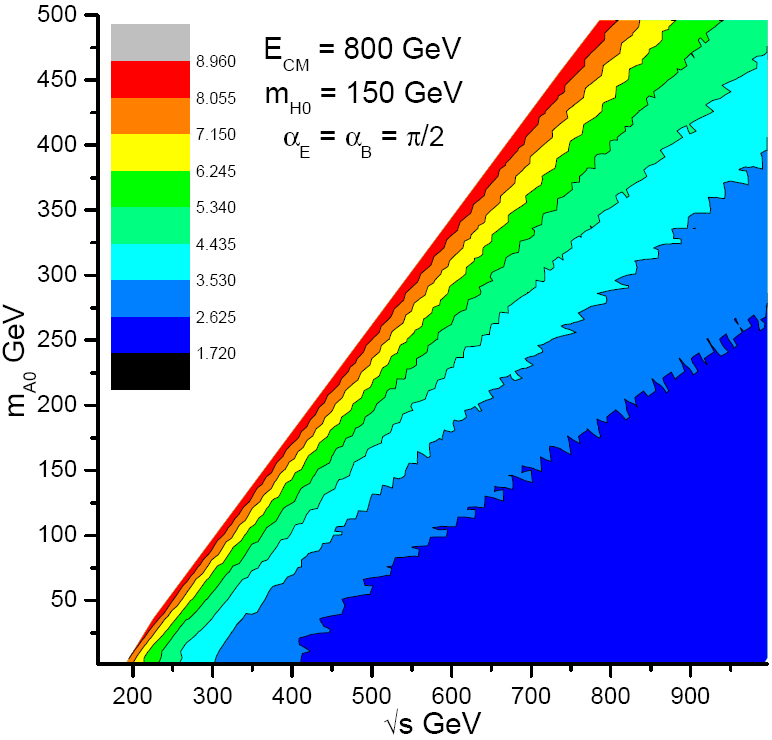}
   \caption{The fractional rate $\sigma_{pol.}/\sigma_{Unpol.}$ on background photon polarized mode $p_{e}$ = 1, $p_{l}$ = -1 and the unpolarized mode $p_{e}$ = 0, $p_{l}$ = 0 is truncated in the line $\sqrt{s}$ = $m_{A^{0}}$+$m_{H^{0}}$. The total cross section depends on the background parameters $\sin^{2}\alpha_{E, B}$. The background field direction is $\alpha_E$ = $\alpha_{B}$ = $\frac{\pi}{2}$.}
   \label{fig:ghasa}
\end{figure}

From the data analysis, Fig.($\hyperref[fig:ghasa]{14}$), the scalar particle mass $m_{H^{0}}$ should be heavier than the mass $m_{A^{0}}$. At the lower incoming photon energy, particle mass is produced on the vacuum by high energy surroundings. In the case, the central energy in low energy range, the fractional power rate is lower distributed on $m_{H^{0}}$ mode than on the $m_{A^{0}}$ mode vreation. The heavier massive particle needs a higher frequency incoming laser beam at the incident. The wave function contains tiny momentum range distribution than light particle. Below the 500 GeV central energy scale, photon power devotion in this event is dense for $H^{0}$ particle, but unconsolidated for $A^{0}$ particle. The probability of $H^{0}$ particle is lower than $A^{0}$ particle, thus, the mass $m_{H^{0}}$ should be massive than $m_{A^{0}}$.

\section{Conclusion}
In the next order collider experiments, the most intention is to search Higgs particle. We have introduced the properties of background field in photon collider technique on Higgs particle creation process in noncommutative spacetime geometry. The interesting point on the $\gamma\gamma\to H^{0}H^{0}$ process is a complete $\theta$ two order deformations. It is complete forbidden by standard model prediction due to bosonic condition and unitarity restriction. Noncommutative $\theta$ deformed spacetime slightly influences our live-world. Lorentz symmetry due to the background magnetic field and electric field violates Lorentz transformation. Photon-photon collider experiments is a best candidate to test triple gauge boson coupling. The process of $\gamma\gamma\to$ massive neutral Higgs particles is a process of violating and/or condensing unitarity. In probing the evidence of destroying or mixing null vector field and lower velocity particle in $Minkowski$ spacetime, $\gamma\gamma\to$ massive neutral particle is a best proof. In this paper, we analyse photon power distribution on the backreact scattering, the fractional rate between polarized and unpolarized pole is associated with the incoming energy and the final particle mass spectrum. The maximum devotion is manifestly dependent on the incoming photon frequency. In the last two section, we compare the $H^{0}H^{0}$ process with $H^{+}H^{-}$ and $A^{0}H^{0}$ one, and taking a constraint on the restriction of $m_{H^{0}}>m_{A^{0}}$. Power law presents that the scalar Higgs creation is vast and numerous by the central energy upper than the final particle creation scale. For the massless charged Higgs $m^{\pm}$, charged $H^{+}$ and $H^{-}$ particles are as the same fields, due to charge distribution depends on its mass range. In $A^{0}H^{0}$ case, the devotion on the fractional differential
cross section in the higher pole area to lower pole one, the fractional rate are all the same consequences.

\acknowledgments{We will thank Professor Chang for useful discuss. This work is also partially discussed with National Science Council of R.O.C. under contact: NSC-95-2112-M-007-059-MY3 and National Tsing Hua University under contact: 97N2309F1.}

\end{document}